\documentclass[preprint,review,12pt]{elsarticle}

\usepackage{graphics}

\usepackage{amssymb}
\usepackage{amsthm}
\usepackage{amsmath}
\usepackage[numbers,sort&compress]{natbib}

\usepackage{float}
\usepackage{graphicx}
\usepackage{tikz}
\usepackage{pgf}
\usepackage{xcolor}
\DeclareUnicodeCharacter{2212}{-}
\usepackage{array}
\newcolumntype{C}[1]{>{\centering\arraybackslash}m{#1}}
\usetikzlibrary{arrows}
\usepackage{threeparttable}
\usepackage{soul}
\usepackage{subfig}
\usepackage{appendix}
\usepackage{lipsum}  

\usepackage{geometry}
\journal{International Journal of Heat and Mass Transfer}

\begin{document}

\begin{frontmatter}



\title{Modeling and Simulation of the Evaporation and Drying of a Two Component Slurry Droplet}


\author{Anurag Bhattacharjee\fnref{fn1}}
\ead{abhattacharjee@wpi.edu}
\fntext[fn1]{Graduate Student}

\author{Aswin Gnanaskandan\corref{cor1}\fnref{fn2}}
\ead{agnanaskandan@wpi.edu}
\cortext[cor1]{Corresponding author}
\fntext[fn2]{Assistant Professor}

\address{Department of Mechanical and Materials Engineering, Worcester Polytechnic Institute, USA}

\begin{abstract} 
In this paper, we present a mathematical model and numerical simulation of the evaporation and drying process of a liquid droplet containing suspended solids. This type of drying is commonly encountered in manufacturing processes such as spray drying and spray pyrolysis, which have applications in industries such as food and pharmaceuticals. The proposed model consists of three stages. In the first stage, we consider the evaporation of the liquid in the presence of solid particles. The second stage involves the formation of a porous crust around a wet core region, with liquid evaporation occurring through the crust layer. Finally, the third stage involves sensible heating of the dry particle to reach ambient temperature. To solve the physical models governing these processes, we use a finite difference method with a moving grid methodology. This allows us to account for the moving interface between the crust and the wet core region of the droplet. In this study, we use a non-uniform temperature model that takes into account the spatial variation of temperature inside the droplet. We also assess the validity of a uniform temperature model. To validate our model, we compare it with experimental data on the drying of a single droplet containing colloidal silica particles. We find that our model agrees well with the experimental results. We rigorously examine assumptions made in the model, such as the shape of the solid particles and the continuum flow of vapor through the porous crust. In addition, we analyze the effects of drying conditions, such as the velocity, temperature, relative humidity, and concentration of solid particles, on the drying rate and the final morphology of the particle. Finally, we develop a regime map that can be used to determine whether the final particle will be solid or hollow, based on the operating conditions.

\end{abstract}


\end{frontmatter}

 \section{Introduction}
 \label{sec:intro}
Evaporation and drying of slurry droplets containing solid particles has several applications in the food~\cite{selvamuthukumaran2019handbook} and pharmaceutical industries~\cite{broadhead1992spray, samborska2022innovations} in processes such as spray drying and spray pyrolysis. In these processes, a wet slurry droplet is dried by exposing it to a hot gas, leading to liquid evaporation and droplet-to-particle conversion. The kinetics of these drying processes is usually controlled by the mass and energy transfer rates at the single-droplet level. Hence, despite inter droplet interactions that are usually present in large-scale drying processes, the study of single-droplet drying, its conversion to a dry particle, and the factors affecting this conversion are of paramount importance, which is the main focus of this study.
\begin{figure}[h!]
	\centering
	\includegraphics[width=0.7\linewidth]{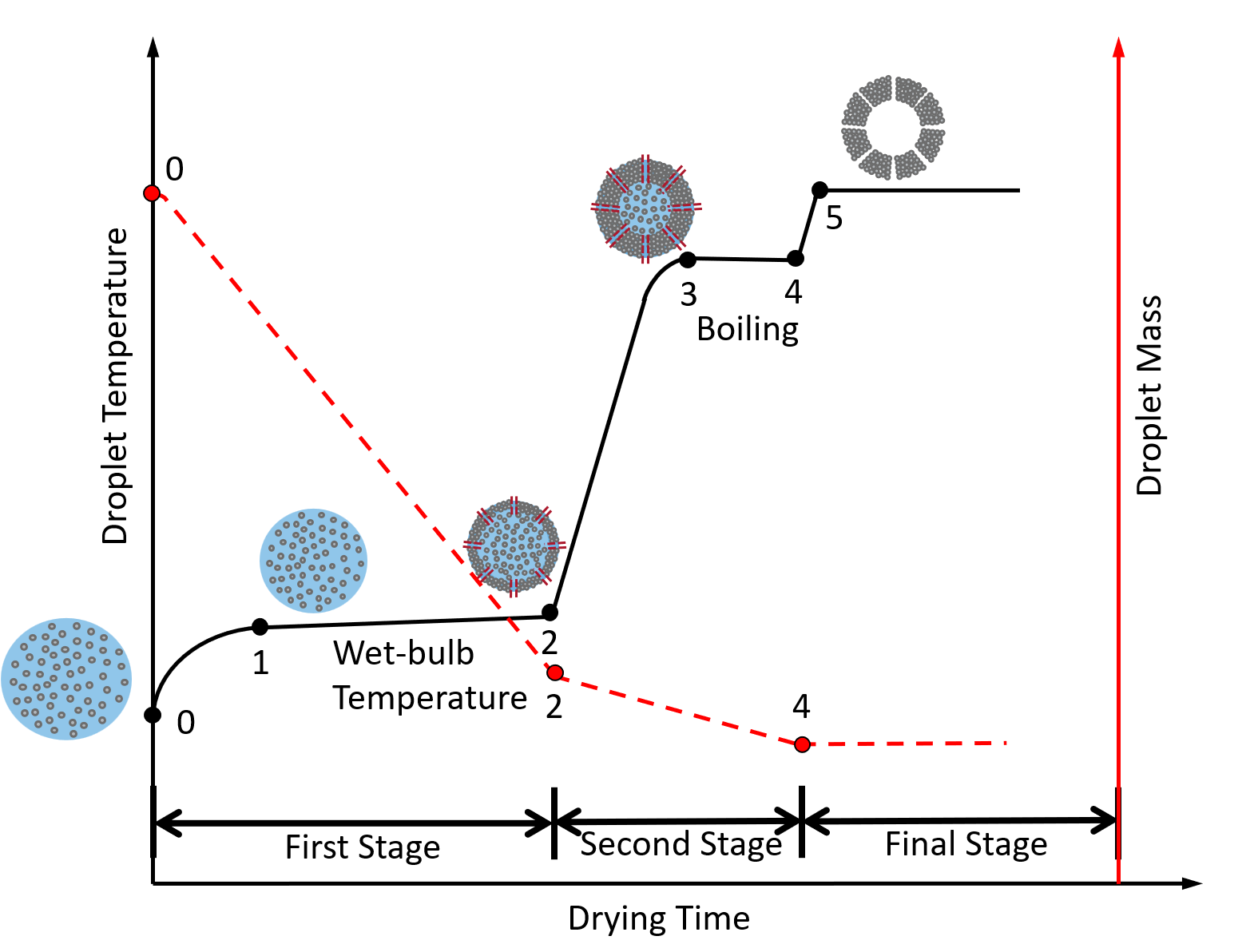}
	\caption{Schematic of the three-stage drying process of a single droplet containing solids showing temperature variation with drying time.}\label{fig:curve}
\end{figure}

The drying process of a single slurry droplet can be divided into three distinct stages, as shown in Figure~\ref{fig:curve}. In the first stage, as the droplet is exposed to the hot drying gas, it undergoes a temperature increase (0-1). This temperature rise continues until the droplet reaches its wet bulb temperature, signaling the start of vaporization and the consequent reduction in droplet size. As evaporation primarily occurs at the droplet's surface, moisture loss leads to the aggregation of solid particles on the surface of the droplet (1-2). This process continues until the solid concentration on the surface attains a critical value, at which point a thin, porous crust is formed on the droplet surface. It is assumed that this crust has sufficient structural integrity to support itself, thus maintaining a constant particle diameter throughout the remaining drying period. This marks the beginning of the second stage, which is characterized by a wet core region within the particle and a porous crust on the surface. In this second stage, the wet core loses moisture from the interface between the wet core and the crust regions. This process leads to an increase in crust thickness and a reduction in the radius of the wet core (2-3) accompanied by an increase in droplet temperature. There is also the possibility of the liquid boiling if the droplet temperature reaches the boiling point of the liquid (3-4). The second stage of drying persists until all moisture has evaporated from the particle. Lastly, the final stage (4-5) represents sensible heating of the particle to ambient temperature.

Several mathematical models have been developed to account for heat and mass transfer during the drying of a single slurry droplet. A mathematical model developed by Parti and Palancz~\cite{parti1974mathematical} was based on an analytical treatment of the drying kinetics of particles. However, the model could not correctly predict the temperature of the final particle.  Ne{\v{s}}i{\'c} and Vodnik~\cite{nevsic1991kinetics} presented a theoretical model for the drying of single droplets containing solids, without taking into account the temperature distribution within the droplets for both drying stages. The model did not account for the effect of Stefan flow on heat transfer from the droplet. The model also showed a strong disagreement on the evolution of the temperature during the second drying stage. Farid~\cite{farid2003new} developed a drying model taking into account the average moisture content of the droplets. The model accounted for the distribution of temperature inside the droplet during the initial heating period, where evaporation was assumed to take place at the wet bulb temperature. The porosity of the crust was not taken into account, which led to discrepancies in the second drying stage. Seydel et al.~\cite{seydel2004experiment,seydel2006modeling} used a one-dimensional population balance approach to model the particle distribution and crystallization within a droplet during drying. The model was validated by performing free-fall experiments using droplets with variations in the initial solid volume fractions and changes in air temperature. Dalmaz et al.~\cite{dalmaz2007heat} also developed a numerical model that included two stages of drying. The model assumed a constant droplet morphology during the drying process. The model also assumed saturation conditions, while accounting for vapor volume fraction on the droplet surface without accounting for the presence of solid particles. The model was validated against the experiments conducted by Ne{\v{s}}i{\'c} and Vodnik~\cite{nevsic1991kinetics} and the simulation results closely matched the experimental data. An inconsistency was observed between the temperature at the center and the surface of the final particle, which is most likely due to the assumption of saturation conditions at the droplet surface.  Handscomb et al.~\cite{handscomb2009new,handscomb2009new1} introduced a model for a single droplet containing a multicomponent mixture in the form of solvent, solute, and solid. A population balance approach was used for the discrete solid phase, a volume-averaged differential mass balance approach was used for the dissolved solute, and the Stefan flow effect was ignored. The temperature of the droplet was assumed to be uniform for both drying stages. They validated the model against experiments conducted on pure water droplet drying, colloidal silica, and sodium sulfate droplets. However, the model underestimated the temperature of the colloidal silica droplets at a drying gas temperature of 101$^\circ$C. Mezhericher et al.~\cite{mezhericher2007theoretical} developed a two-stage drying model with uniform temperature droplets for the first stage and an internal temperature distribution for the second stage. Their model was validated against colloidal silica and skim milk droplet drying experiments and showed good agreement with the experimental data. However, their model could not capture the boiling process for colloidal silica droplet at drying temperature of 178$^\circ$C. In a later study~\cite{mezhericher2008modelling} they also looked at the cracking / breaking of the particles under the influence of the temperature gradient in the crust region. This was achieved by assuming the crust to be a pseudoporous medium and then mathematically modeling the thermal stresses in a particle. Subsequently, Mezhericher et al.~\cite{mezhericher2008heat} developed a model that studied the unsteady characteristics of the second drying stage in great detail. This was achieved by calculating the pressure and vapor fractions inside the crust pores, which are treated as capillary tubes. However, it was also noted that the model was computationally very expensive. Mezhericher et al.~\cite{mezhericher2015multi} also studied droplet-droplet and particle-particle collisions in spray drying, including gas-spray mixing. 


Despite these comprehensive modeling efforts, there exist a few deficiencies in our understanding of this phenomenon, and the objective of this paper is to rectify these knowledge gaps. Current models do not elucidate how the shape of solid particles affects crust formation. However, the saturation concentration on the surface and crust dynamics are influenced by the shape of the particles. In this study, we examine the impact of particle shape on mass and temperature changes by varying their packing factor. Current models also assume that the diffusion of the vapor through the capillary pores follows the continuum assumption. By considering the diffusive Knudsen number, we demonstrate that taking into account non-continuum effects can significantly alter the temperature evolution of the droplet. Several previous studies have employed a uniform temperature model~\cite{parti1974mathematical,nevsic1991kinetics,handscomb2009new,jayanthi1993modeling}, without investigating the conditions under which this assumption holds. To address this gap, we propose a nonuniform temperature model for all three stages and utilize the Biot number as a metric to evaluate the significance of temperature variation in space. Finally, parametric studies conducted in the past~\cite{maevski2010single,eslamian2006modelling} have only elucidated the effect of operating conditions on the evolution of the droplet mass and temperature. In this study, we elucidate the effect of operating conditions on crust formation and also develop a regime map to predict whether the final particle will be solid or hollow. 

The structure of the paper is as follows. We start by presenting the governing equations and a numerical technique for the three stages of droplet drying. Then, we verify the model by comparing it with experimental data and perform a grid independence study to identify the most appropriate grid setup for future simulations. After that, we examine the impact of solid particle shape and assess the assumption of a continuous flow of vapor through the pores. Lastly, we explore the influence of operating conditions on the droplet's morphology and create a regime map to predict whether the particle will be solid or hollow under a specific operating condition.

\section{Physical Model and Numerical Method}\label{}

The governing equations for the three stages of drying are presented \cite{bhattacharjee2023numerical}. The main assumption is that the droplet remains spherical, and hence the equations can be solved in spherical coordinates. The drying medium is assumed to be a binary mixture consisting of dry air and water vapor, and the droplet is assumed to consist of a liquid solvent and solid particles.

\subsection{First Drying Stage}

\begin{figure}[H]
\begin{tikzpicture}[scale=1]
    \draw (-4, 0) node[inner sep=0] {\includegraphics[width=3.1in, trim=0cm 0cm 0cm 0cm, clip]{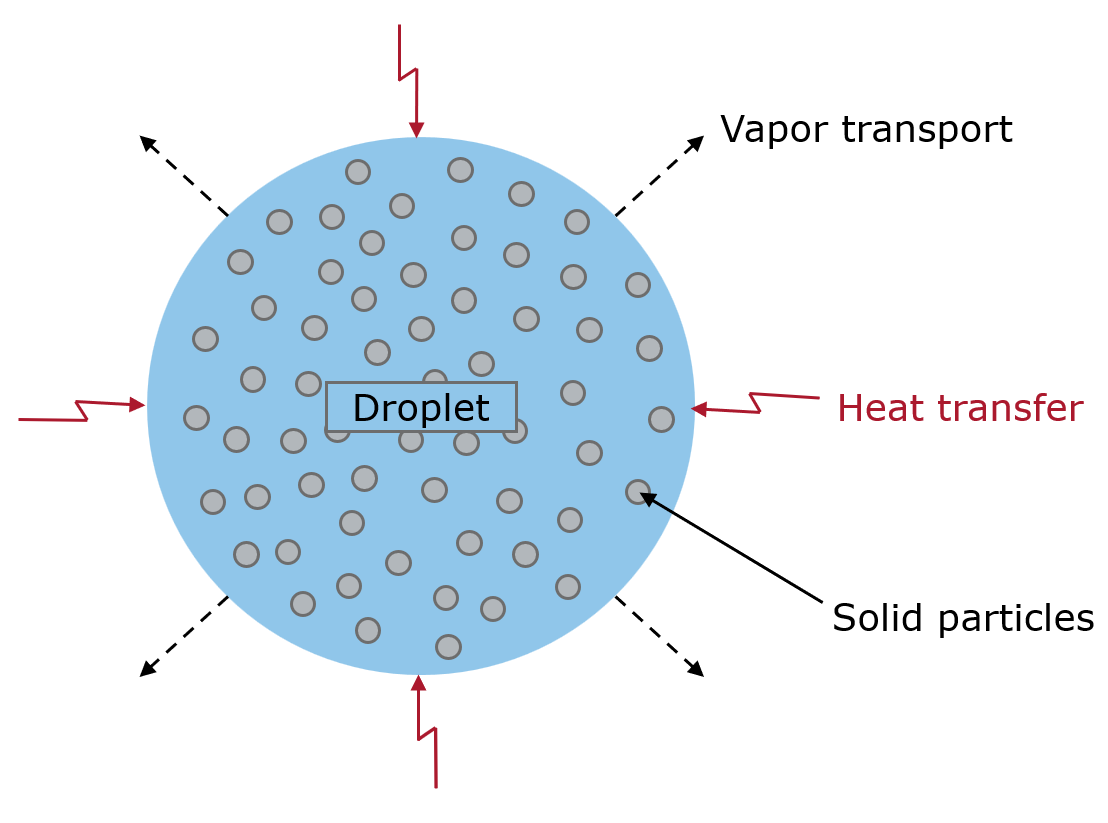}};
    \draw (4, 0) node[inner sep=0] {\includegraphics[width=3.1in, trim=0cm 0cm 0cm 0cm, clip]{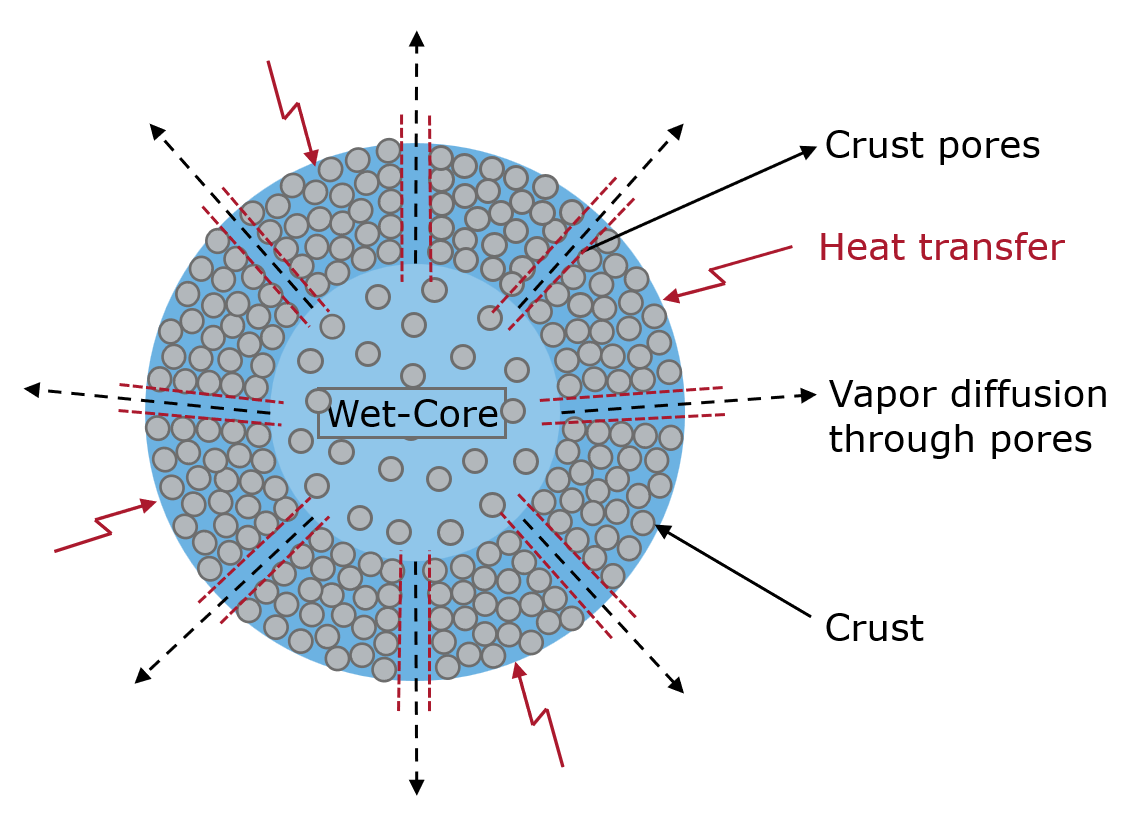}};
    \draw (-7, 3) node[scale=1]{$(a)$};
    \draw (1, 3) node[scale=1]{$(b)$};
\end{tikzpicture}
\caption{Schematic illustrating the physical processes occurring during  (a)  First drying stage and; (b) Second drying stage.}
\label{fig:3}
\end{figure}

Figure \ref{fig:3}(a) shows a schematic of the various heat and mass transport processes that occur in the first stage. Initially, the droplet is assumed to have a uniform temperature and solid concentration. As it is heated, its temperature increases until the wet bulb temperature is reached, after which solvent evaporation begins. Many previous studies~\cite{parti1974mathematical,nevsic1991kinetics,handscomb2009new,mezhericher2007theoretical} assume a uniform droplet temperature, whose validity will be evaluated later in this study. Our model accounts for the spatial and temporal evolution of temperature using the following energy equation~\cite{mezhericher2007theoretical}:
\begin{equation}\label{eqn:energyfirst}
	\rho_d C_{p,d} \frac{\partial T_{d}}{\partial t} = \frac{1}{r^2}\frac{\partial}{\partial r}\left( k_d r^2\frac{\partial T_{d}}{\partial r}   \right),
\end{equation}
where $\rho_d$, $C_{p,d}$, $T_d$, and $k_d$,  denote the density, specific heat, temperature, and thermal conductivity of the droplet, respectively. The boundary conditions to solve this equation are given by:
\begin{equation}\label{eqn:bcfirst}
	\begin{cases}
		\frac{\partial T_d}{\partial r} = 0, & r = 0,\\
		h(T_g - T_d) = k_d\frac{\partial T_d}{\partial r} + h_{fg}\frac{\dot{m_v} }{A_d(t)}, & r = R_d(t),\\
	\end{cases}    
\end{equation}
where $h$, $T_g$, $h_{fg}$, $\dot{m_v}$, $A_d(t)$ and $R_d(t)$ denote the heat transfer coefficient, the temperature of the drying gas, the latent heat of evaporation, the rate of evaporative mass transfer, the droplet area and the droplet radius at time $t$, respectively. The specific heat of the droplet is calculated by:
\begin{equation}
	C_{p,d} = C_{p,l}(1-c) + C_{p,s}c,
\end{equation}
where $C_{p,l}$, $C_{p,s}$, and $c$ denote the specific heat of the liquid, the specific heat of the solid, and the mass concentration of the solid particles, respectively.  The density of the droplet is determined by:
\begin{equation}
	\rho_d = \frac{(1+X)\rho_{d,s}\rho_{d,l}}{\rho_{d,l}+X\rho_{d,s}},
\end{equation}
where $X$, $\rho_{d,s}$ and $\rho_{d,l}$ denote the moisture content of the droplets, the density of the solid particles and the density of the liquid in the droplet, respectively. 

The mass transfer rate of the solvent vapor leaving the droplet surface is given by:
\begin{equation}\label{eqn:masstransfer_first}
	\dot{m_v} = h_D\pi  D_{d}^2(\rho_{v,s}-\rho_{v,g}),
\end{equation}
where $h_D$, $D_d$, $\rho_{v,s}$ and $\rho_{v,g}$ represent the mass transfer coefficient, the diameter of the droplet, the density of vapor at the surface of the droplet and the density of vapor in the drying gas, respectively. Modified Ranz-Marshall equations are used to determine the Nusselt and Sherwood numbers, which are then used to calculate the heat and mass transfer coefficients~\cite{ranz1952evaporation,harpole1981droplet}:
\begin{equation}
	Nu = \left(2 + 0.6 Re^{\frac{1}{2}} Pr^{\frac{1}{3}} \right) (1+B)^{-0.7},
\end{equation}
\begin{equation}
	Sh = \left(2 + 0.6 Re^{\frac{1}{2}} Sc^{\frac{1}{3}} \right) (1+B)^{-0.7},
\end{equation}
where $Re$, $Pr$, and $B$ denote the Reynolds number, Prandtl number, and Spalding mass transfer number. These modified equations are used to account for the Stefan flow that occurs at the droplet surface. The shrinkage rate of the droplet is calculated using~\cite{levi1995mathematical}:
\begin{equation}
	\frac{d(R_d(t))}{dt} = - \frac{1}{\rho_{d,l} 4\pi R_{d}^2 }\dot{m_v}.  
\end{equation}

The solid concentration inside the droplet is obtained using Fick's law of diffusion ~\cite{cussler2009diffusion}:
\begin{equation}
	\frac{\partial C_l}{\partial t} = \frac{1}{r^2}\frac{\partial}{\partial r}\left( r^2 D_{sl}\frac{\partial C_l}{\partial r}      \right),
\end{equation}
The corresponding boundary conditions are given as follows.
\begin{equation}
	\begin{cases}
		\frac{\partial C_l}{\partial r} = 0, & r = 0\\
		-D_{sl}\frac{\partial C_l}{\partial r} + [\rho_l - C_l]\frac{d R_d(t)}{d t} = 0, & r = R_d(t),\\
	\end{cases}    
\end{equation}
where $D_{sl}$ and $C_l$ represent the diffusion coefficient of liquid and liquid mass concentration, respectively. The solid mass concentration, $C_s$ can then be calculated using the following relation:
\begin{equation}
	C_s = \rho_s\left( 1 - \frac{C_l}{\rho_l}   \right).
\end{equation}

\subsection{Second Drying Stage}

Figure~\ref{fig:3}(b) shows a schematic of the second stage of droplet drying. Once the solid concentration on the droplet's surface reaches a critical value, it is assumed that a thin porous crust will form, marking the beginning of the second stage. This results in the transformation of the droplet into a wet particle, consisting of an inner wet core and an outer dry crust. It is assumed that the particle's outer diameter will remain constant during the second stage of drying. As the drying process progresses, the inner wet core continues to lose moisture, which escapes through the pores of the crust. The amount of vapor leaving the wet particle can be estimated using Darcy's law. As the wet core continues to lose moisture, the radius of the interface decreases, while the thickness of the crust increases.

The governing equations of the droplet are now divided into two zones, one for the wet core and one for the dry crust. The energy balance for the crust region is given by~\cite{mezhericher2007theoretical}:
\begin{equation}\label{eqn:crust}
	\rho_{cr} C_{p,cr}  \frac{\partial T_{cr}}{\partial t} = \frac{1}{r^2}\frac{\partial}{\partial r}\left( k_{cr} r^2\frac{\partial T_{cr}}{\partial r}      \right),
\end{equation}
where $\rho_{cr}$, $C_{p,cr}$, $T_{cr}$, $k_{cr}$, denote the density, specific heat, temperature, and thermal conductivity of the crust, respectively. The boundary conditions for the crust region are given by:
\begin{equation}\label{eqn:bcsecondcr}
	\begin{cases}
		k_{cr}\frac{\partial T_{cr}}{\partial r} = k_{wc}\frac{\partial T_{wc}}{\partial r} + h_{fg} \frac{\dot{m_v}}{A_i}, & r = R_i(t),\\
		T_{wc} = T_{cr}, & r = R_i(t),\\
		h(T_g - T_{cr}) = k_{cr}\frac{\partial T_{cr}}{\partial r} , & r = R_p,\\
	\end{cases}    
\end{equation}
where $T_{wc}$, $k_{wc}$, $R_i$, and $R_p$ denote the temperature and thermal conductivity of the wet core, the radius of the wet core - dry crust interface and the radius of the wet particle, respectively. Similarly, the energy balance for the wet core is given by:
\begin{equation}\label{eqn:wetcore}
	\rho_{wc} C_{p,wc}  \frac{\partial T_{wc}}{\partial t} = \frac{1}{r^2}\frac{\partial}{\partial r}\left( k_{wc} r^2\frac{\partial T_{wc}}{\partial r}      \right).
\end{equation}
The corresponding boundary conditions for the wet core are given by:
\begin{equation}\label{eqn:bcsecondwc}
	\begin{cases}
		\frac{\partial T_{wc}}{\partial r} = 0, & r = 0,\\
		T_{wc} = T_{cr}, & r = R_i(t),\\
		k_{cr}\frac{\partial T_{cr}}{\partial r} = k_{wc}\frac{\partial T_{wc}}{\partial r} + h_{fg} \frac{\dot{m_v}}{A_i}, & r = R_i(t),\\
	\end{cases}    
\end{equation}
where $\rho_{wc}$ and $C_{p,wc}$ denote the density and specific heat of the wet core. The mass of the wet particle can be estimated by:
\begin{equation}
	m_p = \frac{m_{d,0}}{1+X_0} \left ( 1-\frac{\rho_{wc,l}}{\rho_{wc,s}} \right) + \frac{4}{3}\pi \rho_{wc,l} \left( \varepsilon R_i^3 + (1-\varepsilon)R_p^3  \right),
\end{equation}
where $m_p$, $m_{d,0}$, $X_0$, $\rho_{wc,l}$, $\rho_{wc,s}$ and $\varepsilon$ stand for the mass of the particle, the initial mass of droplet, initial moisture content, the density of the liquid in the wet core, density of the solid in wet core and porosity of the crust region, respectively. The mass of dry crust can be obtained using:
\begin{equation}
	m_{cr} = \frac{4}{3}\pi (1 - \varepsilon)\rho_{cr,s}\left( R_p^3 - R_i^3 \right).
\end{equation}
The crust acts as a porous medium that allows flow of vapor from the interface. This effect is taken into account through Darcy's law, and the following expression is used for the mass transfer rate of the vapor from the wet particle~\cite{abuaf1986drying}:
\begin{equation}
	\dot{m_v} = -\frac{8\pi \epsilon^\beta D_{v,cr} M_l P_g}{R(T_{cr,s}+T_{wc,s})} \frac{R_p R_i}{R_p-R_i} \\ \times \ln{\left(\frac{P_g-P_{v,i}}{P_g - \left( \frac{R}{4\pi M_l h_D R_p^2}\dot{m_v} + \frac{P_{v,g}}{T_g} \right )T_{cr,s}}\right)},
\end{equation}
where $\beta$, $D_{v,cr}$, $M_l$, $R$, $T_{cr,s}$, $T_{wc,s}$, $P_g$, $P_{v,i}$ and$P_{v,g}$ represent the power coefficient, coefficient of vapor diffusion in crust pores, molar mass of liquid, universal gas constant, temperature at crust surface, temperature at crust-wet core interface,pressure of the drying gas, vapor pressure at the interface and vapor pressure of the surrounding gas respectively. $D_{v,cr}$ is assumed to be equal to the coefficient of vapor diffusion $D_v$ at standard atmospheric pressure at droplet surface and is calculated as~\cite{grigoriev1988thermal}:
\begin{equation}\label{ficksdiff}
	D_v = 3.564.10^{-10}(T_{d,s} + T_g)^{1.75},
\end{equation}
where $T_{d,s}$ and $T_g$ are the temperature at the surface of the droplet and drying gas. For the wet particle, this value is computed on the surface of the crust. The rate of change in the radius of the interface is then given by~\cite{levi1995mathematical}:

\begin{equation}
	\frac{d R_i(t)}{dt} = -\frac{1}{\varepsilon \rho_{wc}4 \pi R_i^2}\dot{m_v}.
\end{equation}

\subsection{Final Drying Stage}

The final stage begins when all the moisture has evaporated, transforming the droplet into a dry particle. Once the dry particle is formed, the droplet mass ceases to decrease any further, while the temperature rapidly rises until it is in equilibrium with the drying gas. The dry core beneath the crust no longer undergoes shrinkage and is instead filled with air. Spatailly non-uniform equations are used to calculate the temperatures of the crust and dry core. The governing equations for the crust and the core are same as equations~\ref{eqn:crust} and~\ref{eqn:wetcore}. The boundary conditions for the crust no longer includes heat loss due to evaporation and the crust gains heat solely through convection. The interface radius is unchanging and does not depend on time. The boundary conditions for the crust are given by:
\begin{equation}\label{eqn:bcsfinalcr}
	\begin{cases}
		k_{cr}\frac{\partial T_{cr}}{\partial r} = k_{dc}\frac{\partial T_{dc}}{\partial r} , & r = R_i,\\
		T_{dc} = T_{cr}, & r = R_i,\\
		h(T_g - T_{cr}) = k_{cr}\frac{\partial T_{cr}}{\partial r} , & r = R_p,\\
	\end{cases}    
\end{equation}
where $T_{dc}$ and $R_i$ denote the dry core temperature and constant interface radius, respectively. The corresponding boundary conditions for the dry core are given by:
\begin{equation}\label{eqn:bcfinaldc}
	\begin{cases}
		\frac{\partial T_{dc}}{\partial r} = 0, & r = 0,\\
		T_{dc} = T_{cr}, & r = R_i,\\
		k_{cr}\frac{\partial T_{cr}}{\partial r} = k_{dc}\frac{\partial T_{dc}}{\partial r}, & r = R_i.\\
	\end{cases}    
\end{equation}

\subsection{Numerical Method}
\begin{figure}[h!]
	\centering\includegraphics[width=0.85\linewidth]{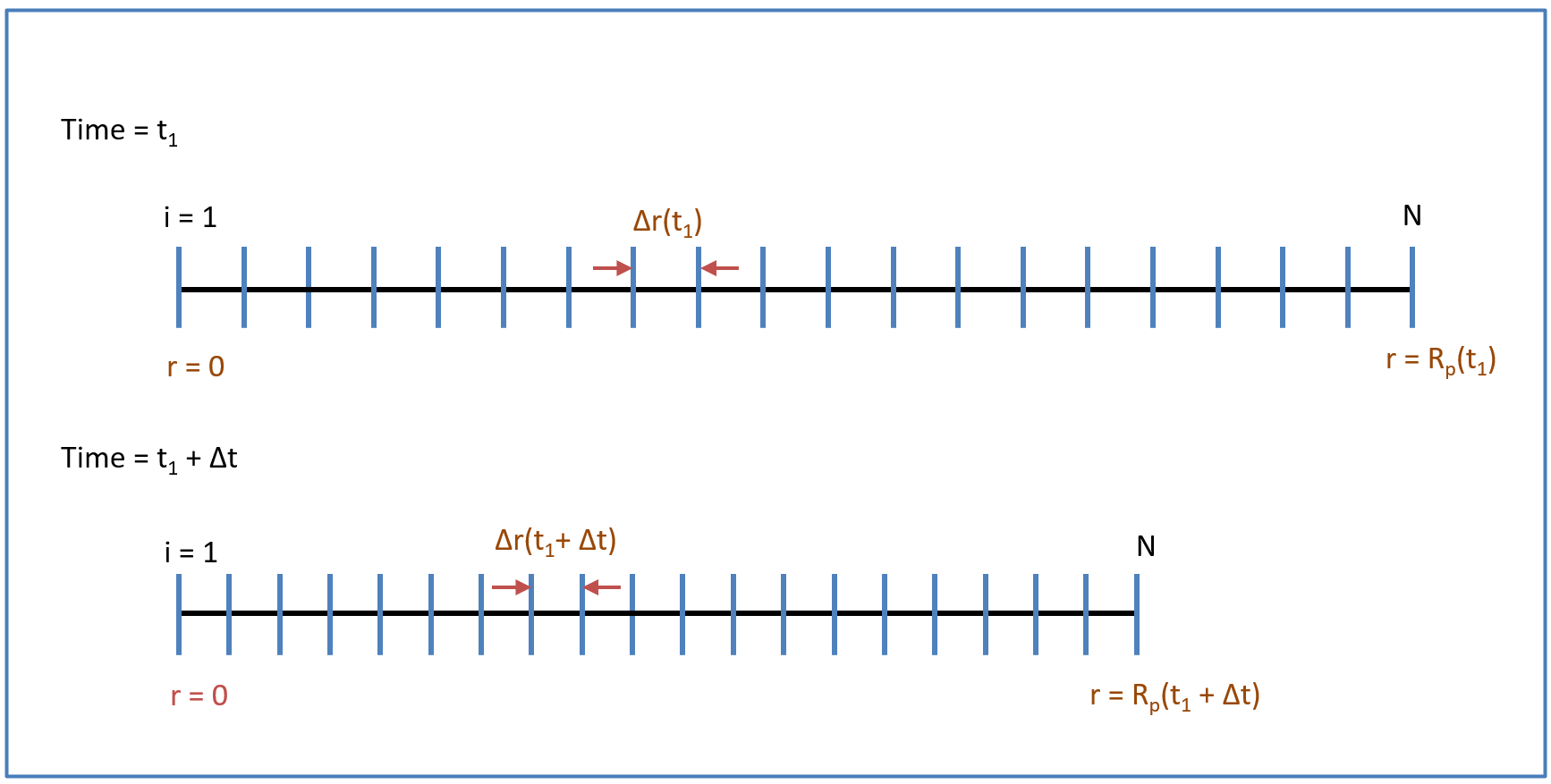}
	\caption{Schematic of the grid configuration for the first stage of drying.}\label{fig:grid_first}
\end{figure}

\begin{figure}[h!]
	\centering\includegraphics[width=0.85\linewidth]{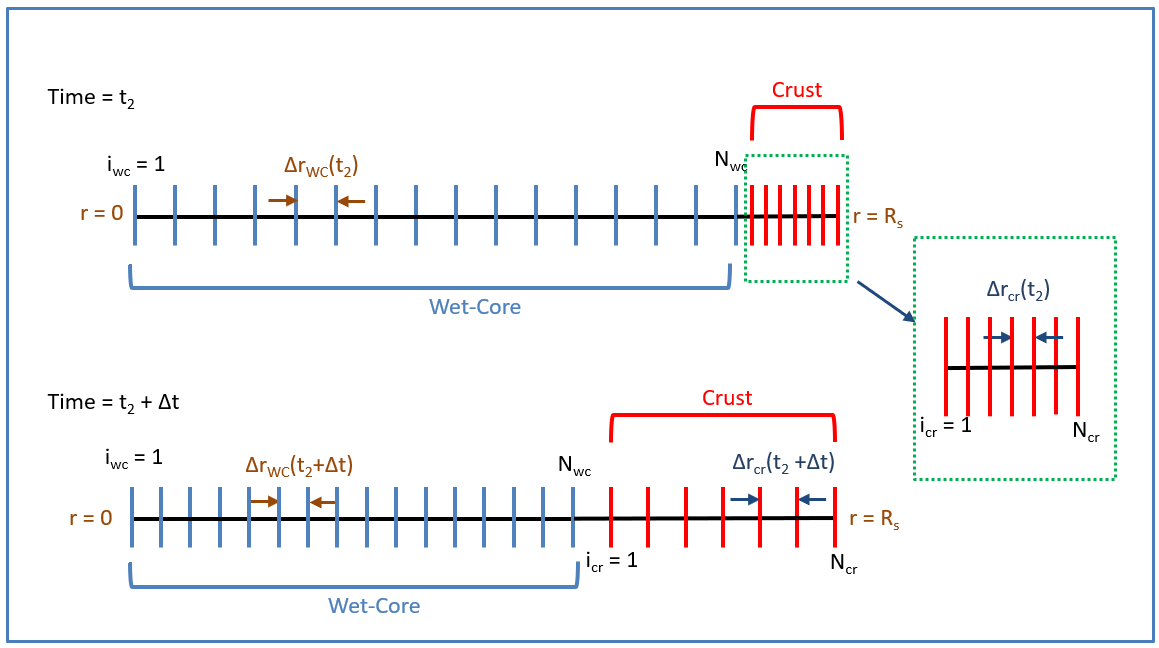}
	\caption{Schematic of the grid configuration for the second stage of drying.}\label{fig:grid_second}
\end{figure}

A moving boundary method is used to determine the mesh in the first and second stages of drying. The number of grid points remains constant as the boundary moves, resulting in a decrease in grid size with each time step as the droplet shrinks. Figure~\ref{fig:grid_first} illustrates that the number of grid points for two different times, $t_1$ and $t_2+\Delta t$, is denoted as N. However, the cell size changes from $\Delta r(t_{1})$ to $\Delta r(t_{1} + \Delta t)$, where $\Delta r(t_{1})$ is larger than $\Delta r(t_{1} + \Delta t)$. 
In the second stage, we begin with a thin crust and a relatively larger wet core. As the drying process continues, the crust becomes thicker while the wet core decreases in size. The number of cells in the crust and wet core remains constant over time. Figure~\ref{fig:grid_second} illustrates that the number of grid points in the crust, $N_{cr}$, and the wet core, $N_{wc}$, remains unchanged at time instances $t_2$ and $t_2+\Delta t$. The crust expands from having a mesh size of $\Delta r_{cr}(t_{2})$ to $\Delta r_{cr}(t_{2} + \Delta t)$, while the wet core contracts, resulting in a reduction in mesh size from $\Delta r_{wc}(t_{2})$ to $\Delta r_{wc}(t_{2} + \Delta t)$.
The governing equations for the mass and energy balance are numerically integrated using a first-order Euler method in time and a second-order central difference method in space.

\section{Results and Discussion}\label{}

\subsection{Model Validation and Description of the Drying Process}
\begin{figure}[H]
\begin{tikzpicture}[scale=1]
    \draw (-4, 0) node[inner sep=0] {\includegraphics[width=3.3in, trim=0cm 0cm 0cm 0cm, clip]{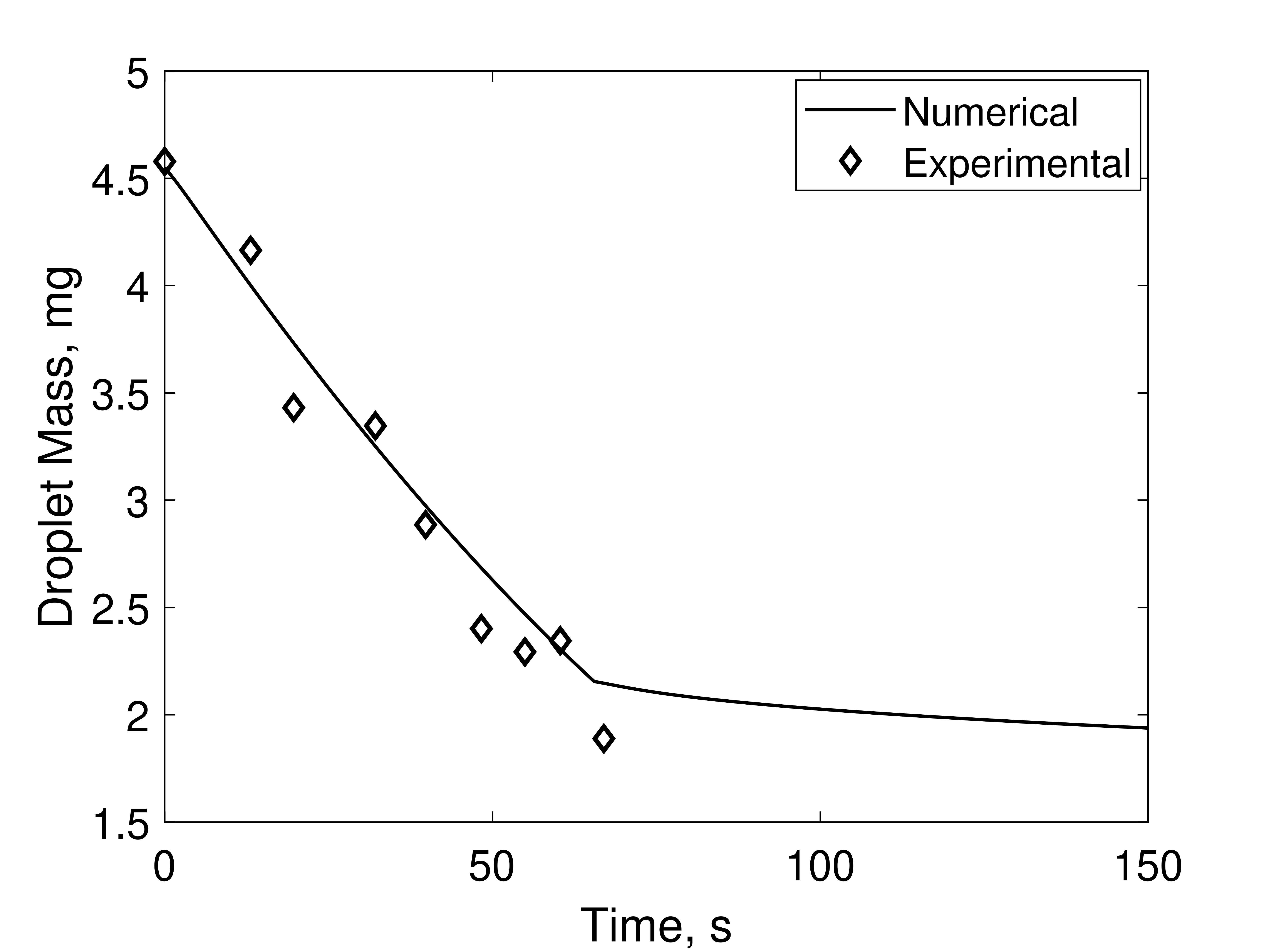}};
    \draw (4, 0) node[inner sep=0] {\includegraphics[width=3.3in, trim=0cm 0cm 0cm 0cm, clip]{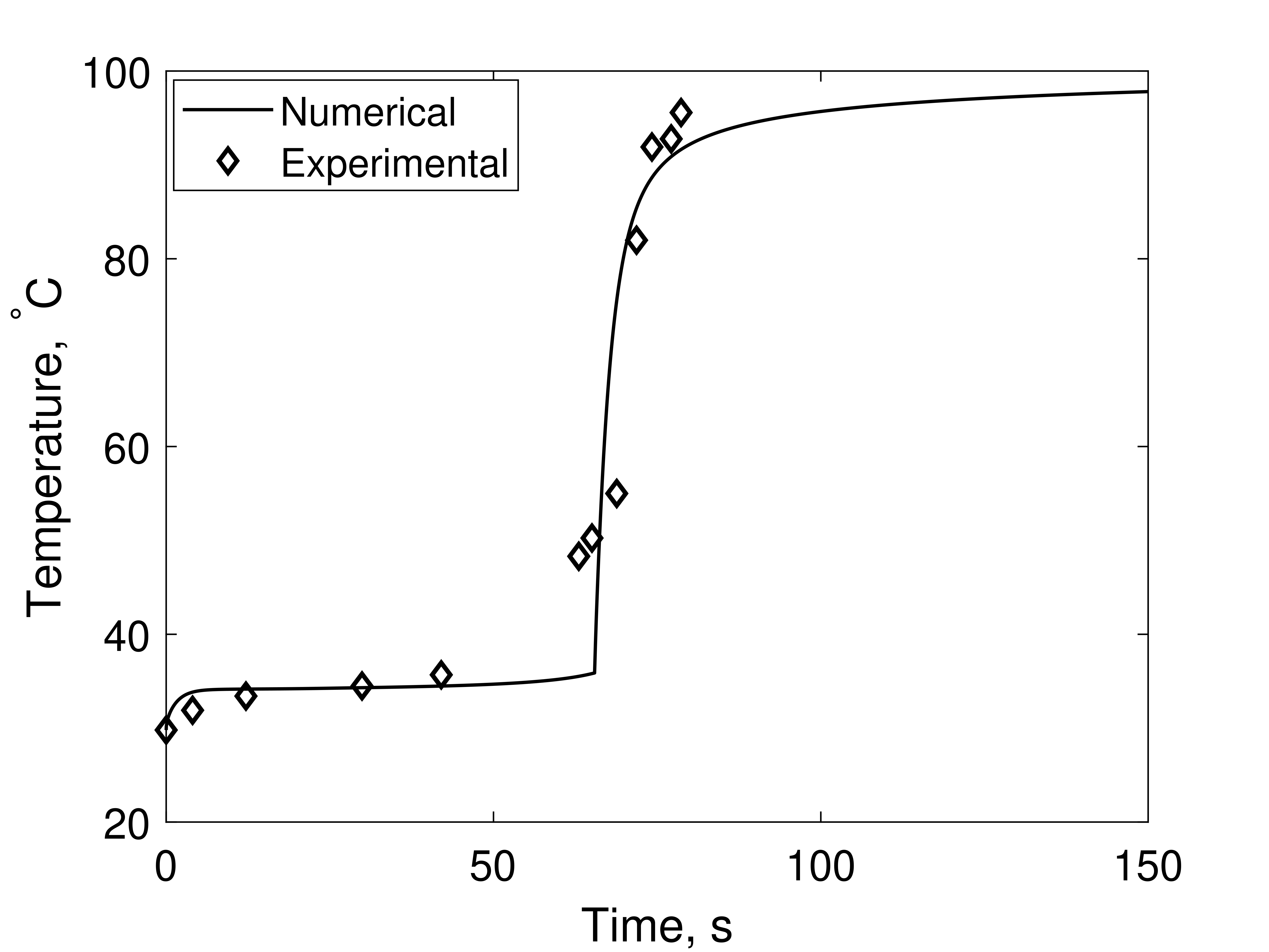}};
    \draw (-7, 3.2) node[scale=1]{$(a)$};
    \draw (1, 3.2) node[scale=1]{$(b)$};
\end{tikzpicture}
\caption{Comparison of numerical and experimental results for colloidal silica droplet drying at an ambient temperature of 101 $^\circ$C showing the evolution (a)  Mass and (b) Temperature.}
\label{fig:silica}
\end{figure}

The proposed model is validated by comparison with the experimental data of Ne{\v{s}}i{'c} and Vodnik~\cite{nevsic1991kinetics} for the drying of colloidal silica particles in water under two different drying conditions. The first experiment involved drying a colloidal silica droplet in air maintained at a relative humidity of 0.4 \%, moving at a speed of 1.73 $ms^{-1}$ relative to the droplet, and an ambient temperature of 101 $^\circ$C. The initial droplet size was 0.98 mm, the initial droplet temperature was 29.8 $^\circ$C, and the initial solid mass fraction was 0.3. Figures \ref{fig:silica}(a) and (b) depict the evolution of the droplet mass and temperature, respectively, which shows good agreement with the experimental data.  
\begin{figure}[H]
\begin{tikzpicture}[scale=1]
    \draw (-4, 0) node[inner sep=0] {\includegraphics[width=3.3in, trim=0cm 0cm 0cm 0cm, clip]{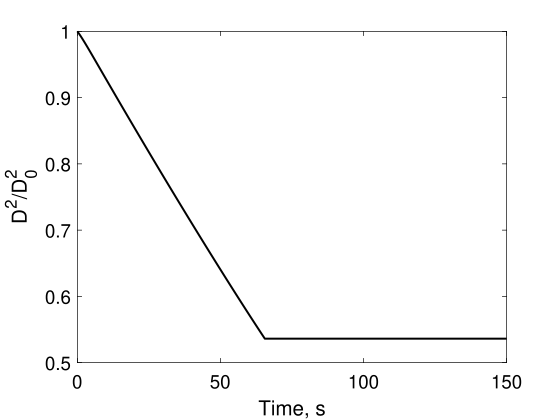}};
    \draw (4, 0) node[inner sep=0] {\includegraphics[width=3.3in, trim=0cm 0cm 0cm 0cm, clip]{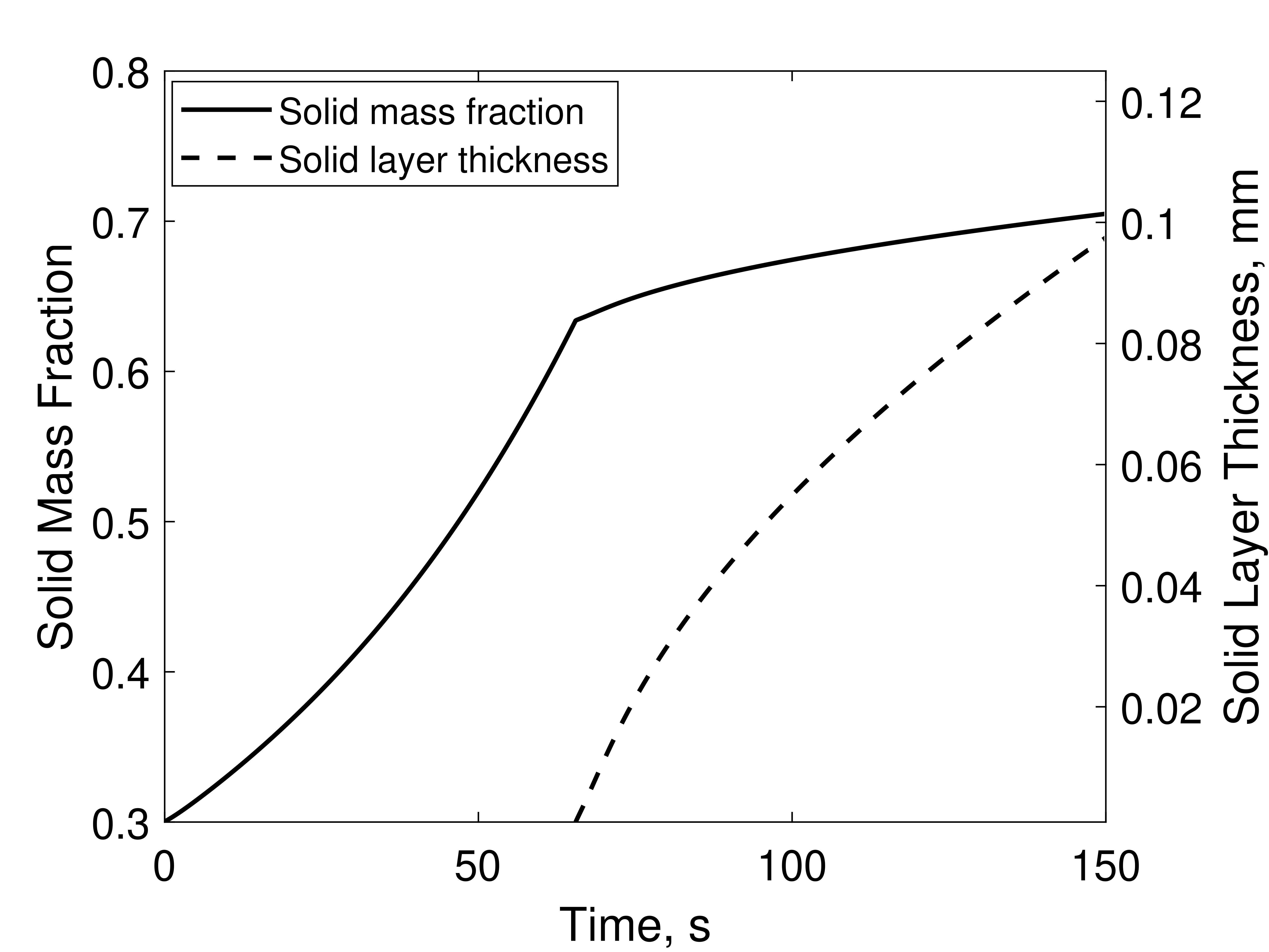}};
    \draw (-7, 3.2) node[scale=1]{$(a)$};
    \draw (1, 3.2) node[scale=1]{$(b)$};
\end{tikzpicture}
\caption{Numerical simulations of the colloidal silica droplet drying at an ambient temperature of 101 $^\circ$C   showing the evolution of (a)  droplet diameter and (b) overall solid mass fraction and crust thickness. }
\label{fig:d2}
\end{figure}
Figures \ref{fig:d2}(a) and (b) illustrate the changes in droplet diameter, overall solid mass fraction, and thickness of the solid crust over time. Initially, the droplet undergoes a drying process, indicated by changes in mass and radius, once it reaches the wet bulb temperature of 33.34$^\circ$ C. Throughout this stage, the temperature remains relatively constant due to the utilization of heat as latent heat. The diameter of the droplet in this stage follows the well-known $D^{2}$ law~\cite{law2010combustion}. As moisture is lost, the solid mass fraction on the droplet's surface increases. When the solid volume fraction reaches a critical value of 0.74, determined by the particle packing efficiency, a thin porous crust of a predetermined thickness forms instantaneously. The impact of the initial crust thickness on the subsequent evolution is discussed in Section~\ref{sec:init_thick}. The crust acts as a barrier, preventing further shrinkage of the droplet and maintaining a constant outer diameter. During this stage, the droplet consists of a solid crust and a wet-core, with the crust being purely solid and the wet-core containing both solid and liquid. As evaporation continues, the wet-core loses moisture through the pores of the crust, resulting in an increase in the solid mass fraction. The temperature gradually rises during this stage, along with the thickness of the crust, as shown in Figures \ref{fig:silica} and \ref{fig:d2}, respectively.

\begin{figure}[H]
\begin{tikzpicture}[scale=1]
    \draw (-4, 0) node[inner sep=0] {\includegraphics[width=3.3in, trim=0cm 0cm 0cm 0cm, clip]{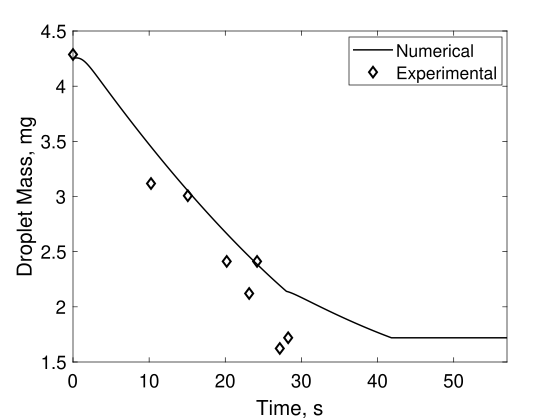}};
    \draw (4, 0) node[inner sep=0] {\includegraphics[width=3.3in, trim=0cm 0cm 0cm 0cm, clip]{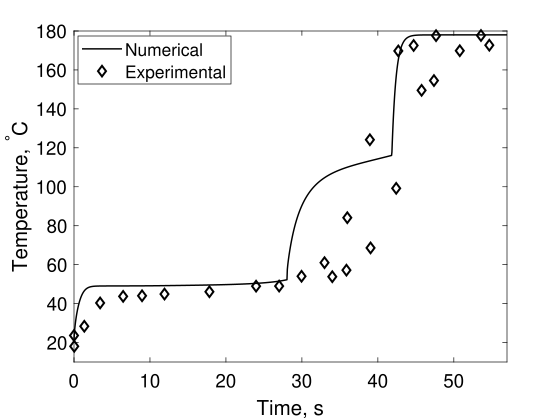}};
    \draw (-7, 3.2) node[scale=1]{$(a)$};
    \draw (1, 3.2) node[scale=1]{$(b)$};
\end{tikzpicture}
\caption{Comparison of numerical and experimental results for the colloidal silica droplet drying at an ambient temperature of 178 $^\circ$C showing the evolution of (a)  Mass and (b) Temperature.}
\label{fig:silica_178}
\end{figure}
In the second experiment, a colloidal silica droplet was dried using a drying gas temperature of 178 $^\circ$C, which is higher than the boiling point of water. The initial radius of the droplet was 0.96 mm and its temperature was 17 $^\circ$C. The drying gas velocity was 2.5 $ms^{-1}$~\cite{handscomb2010simulating}. Figures~\ref{fig:silica_178}(a) and (b) depict the mass and temperature changes observed in both the simulations and experiments, showing a reasonable level of agreement. It is important to note that the temperature rises once the droplet's temperature surpasses 100 $^\circ$C, indicating the onset of boiling, as illustrated in Figure~\ref{fig:silica_178}(b). Once all the moisture has evaporated through boiling, the droplet's mass remains constant. The temperature of the dry particle continues to rise until it reaches the equilibrium with the drying gas temperature of 178 $^\circ$C.

\subsection{Grid Independence Study}

\begin{figure}[H]
\begin{tikzpicture}[scale=1]
    \draw (-4, 0) node[inner sep=0] {\includegraphics[width=3.3in, trim=0cm 0cm 0cm 0cm, clip]{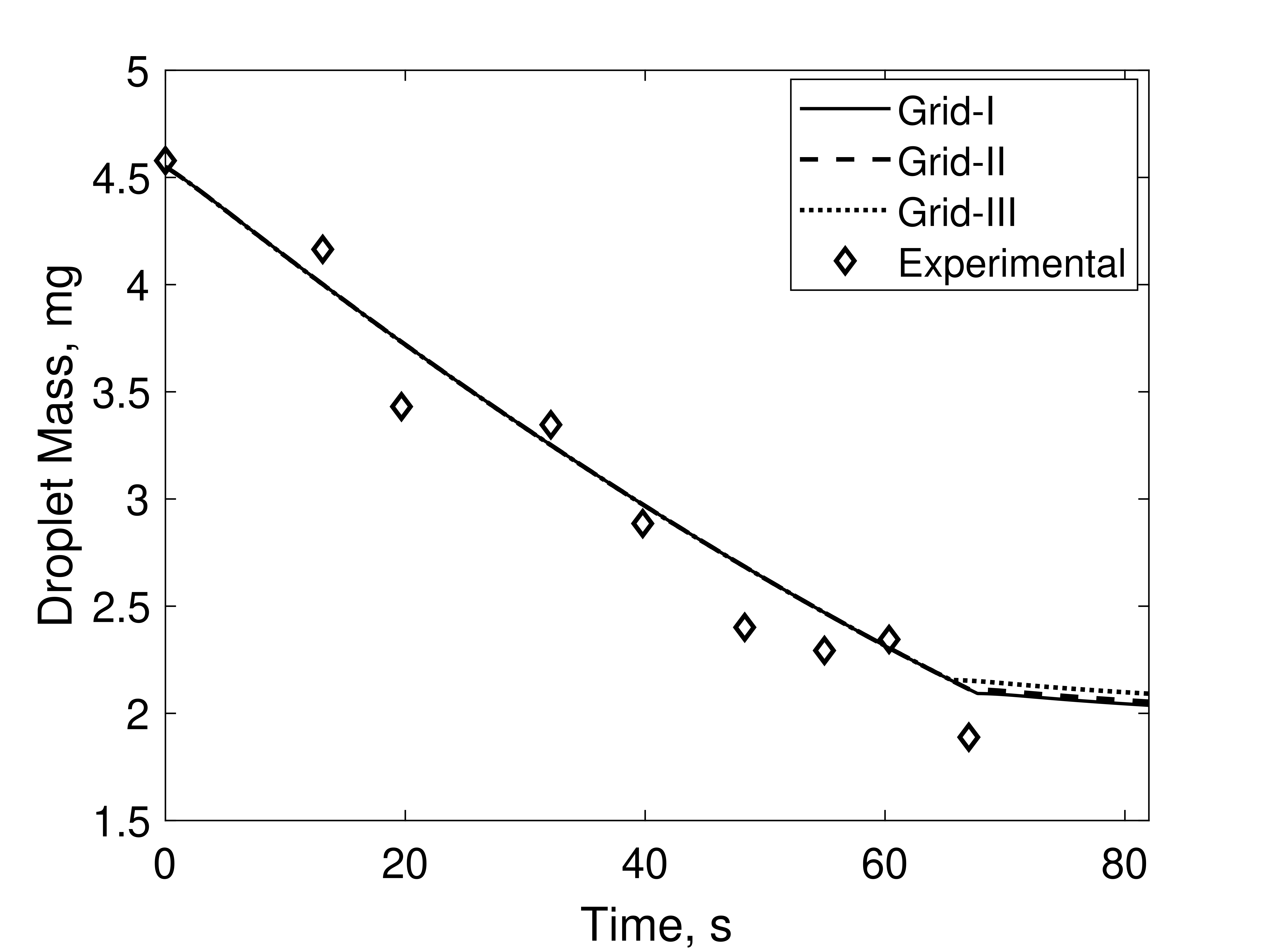}};
    \draw (4, 0) node[inner sep=0] {\includegraphics[width=3.3in, trim=0cm 0cm 0cm 0cm, clip]{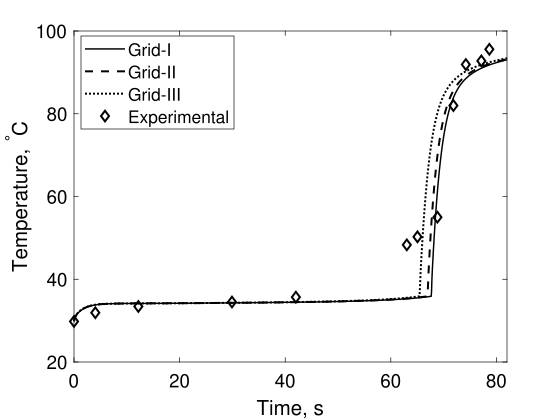}};
    \draw (-7, 3.2) node[scale=1]{$(a)$};
    \draw (1, 3.2) node[scale=1]{$(b)$};
\end{tikzpicture}
\caption{Comparison of numerical results for three different grids with experimental results showing the evolution (a)  Mass and (b) Temperature.}
\label{fig:grid}
\end{figure}

Figures ~\ref{fig:grid}(a) and (b) illustrate the effect of three levels of grid on the droplet mass and temperature evolution. The simulations presented here correspond to the first experiment of Ne{\v{s}}i{\'c} and Vodnik~\cite{nevsic1991kinetics} at a drying gas temperature of 101 $^\circ$C.  These calculations were carried out with the grid characteristics specified in Table ~\ref{tab:grid}.
\begin{table*}[h!]
\caption{Grid characteristics used to conduct the grid independence study.}\label{tab:grid}
\vspace{0.3cm}
\centering{
\begin{tabular*}{1\textwidth}{@{\hspace*{1em}}@{\extracolsep{\fill}}ccccc@{\hspace*{1em}}}
\hline
\multicolumn{1}{c}{\textbf{Drying Stage}} &
\multicolumn{1}{c}{\textbf{Region}} &
\multicolumn{1}{|c}{\textbf{}}&
\multicolumn{1}{c}{\textbf{Number of cells}}&
\multicolumn{1}{c|}{\textbf{}}\\
&&\multicolumn{1}{|c}{\textbf{Grid-I}}&
\multicolumn{1}{c}{\textbf{Grid-II}} &
\multicolumn{1}{c|}{\textbf{Grid-III}}\\
\hline
First &Droplet&80&40&20\\
\hline
Second &Crust&20&10&5\\
&Wet-Core&80&40&20\\
\hline
Final &Crust&20&10&5\\
&Dry-Core&80&40&20\\
\hline
\end{tabular*}
}
\end{table*}
 Simulations using grid I and grid II yield results that exhibit closer alignment with the experimental data. Furthermore, the disparity between the results for these two grids-I and II is negligible, as can be seen in Figure~\ref{fig:grid}. Therefore, grid II is chosen for the subsequent calculations, as the use of grid II, with moderate cell count, saves considerable computational resources.

\subsection{Effect of Model Parameters and Modeling Assumptions}
In this section, our objective is to elucidate the impact of certain modeling parameters and assumptions that have not been adequately addressed in previous research. First, we assess the influence of the thickness of the crust that is assumed to form at the onset of the second stage. Subsequently, we investigate the effect of the shape of the solid particles. Here, we consider the impact of different particle shapes by introducing a packing factor and determining the critical solid fraction for each shape. Furthermore, we examine the presence of non-continuum effects on the vapor transport through the pores of the crust and analyze how the porosity of the crust influences the mass and temperature evolution. Finally, we study the validity of the spatially uniform temperature assumption that has been assumed in some of the previous modeling studies.

\subsubsection{Effect of Initial Crust Thickness}\label{sec:init_thick}

\begin{figure}[H]
\begin{tikzpicture}[scale=1]
    \draw (-4, 0) node[inner sep=0] {\includegraphics[width=3.3in, trim=0cm 0cm 0cm 0cm, clip]{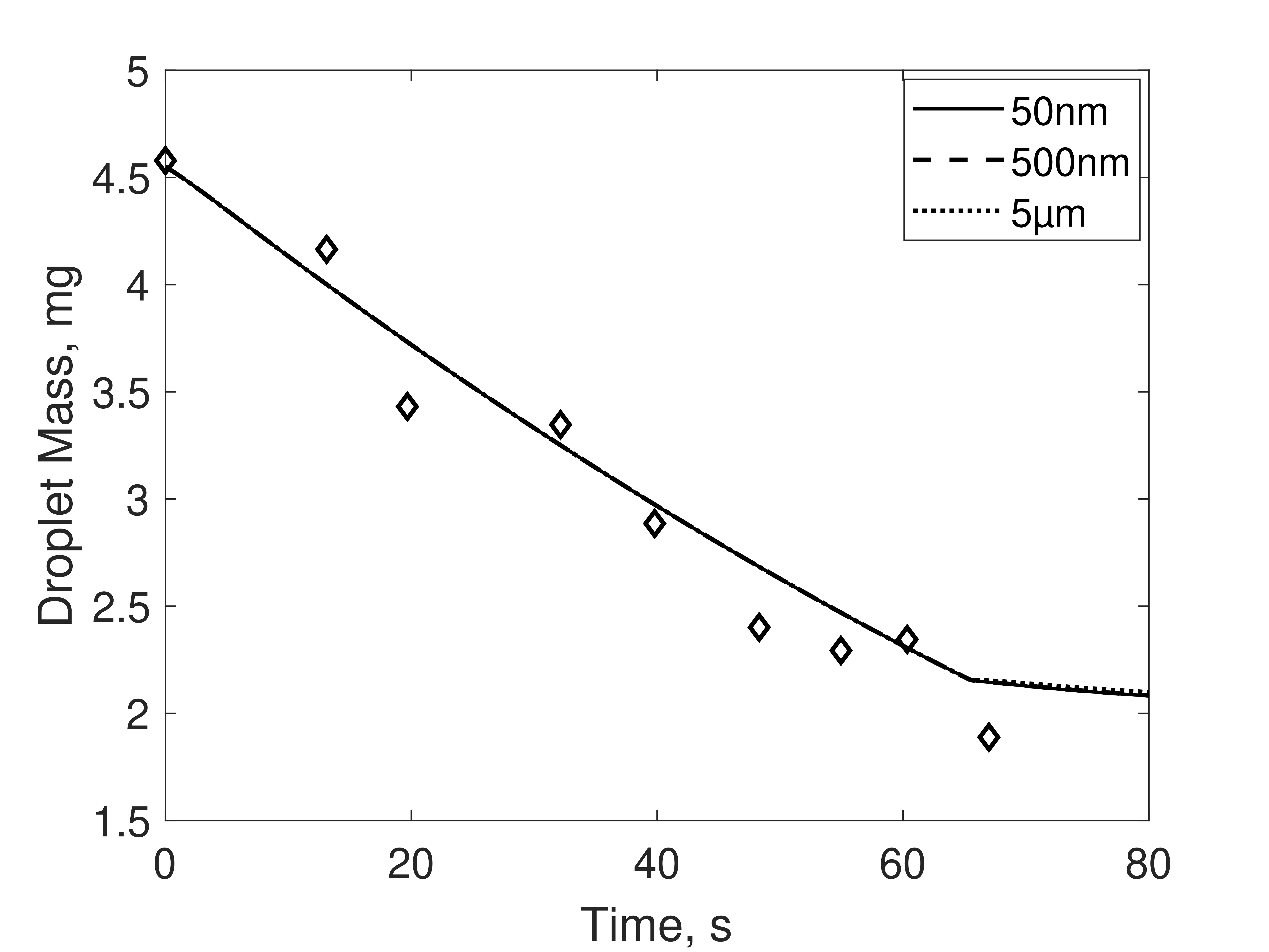}};
    \draw (4, 0) node[inner sep=0] {\includegraphics[width=3.3in, trim=0cm 0cm 0cm 0cm, clip]{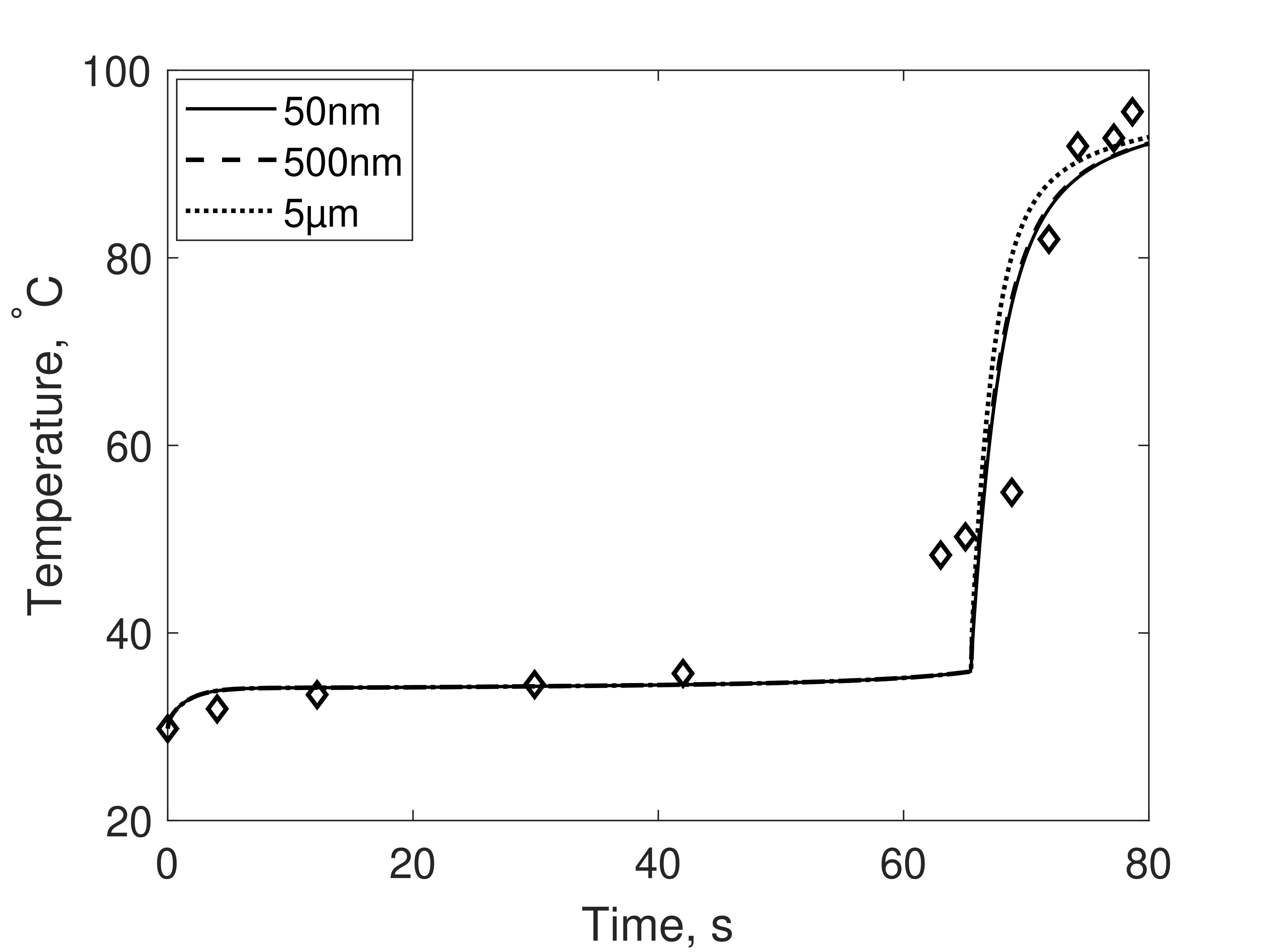}};
    \draw (-7, 3.2) node[scale=1]{$(a)$};
    \draw (1, 3.2) node[scale=1]{$(b)$};
\end{tikzpicture}
\caption{Effect of initial crust thickness on the evolution of (a)  Mass and (b) Temperature.}
\label{fig:init_thick}
\end{figure}

In this section, we investigate the influence of the initial crust thickness assumed to form at the beginning of the second stage. We consider three different initial crust thicknesses: 50$nm$, 500$nm$, and 5$\mu m$, corresponding to a width of 3, 30 and 300 layers of particles. Figure~\ref{fig:init_thick}(a) demonstrates that the variation in the initial crust thickness has minimal effect on the moisture removal rate, as the pore sizes remain unchanged. However, there is a slight difference in the temperature evolution of the crust between an initial thickness of 5$\mu m$ and the other thicknesses, as depicted in Figure~\ref{fig:init_thick}(b). This discrepancy can be attributed to the non-uniform temperature distribution across a thicker crust.

A crust thickness of 500$nm$ is initially chosen for the validation study to balance computational efficiency and accuracy. However, it is important to acknowledge that using an extremely small crust thickness would demand significant computational resources without providing a substantial improvement in temperature evolution accuracy. Hence, for further parametric investigations, a crust thickness of 5$\mu m$ is adopted.




\subsubsection{Effect Of Solid Particle Shape}\label{sec:shape}

The first stage is assumed to end when the solid particles on the surface of the droplet reach a critical volume fraction such that they are sufficiently packed and cannot be compacted any more. Most previous studies~\cite{mezhericher2007theoretical,mezhericher2008heat} have assumed these particles to be spherical in shape. However, depending on the manufacturing process, the shapes of the solid particles can be spherical, tetrahedral, cylindrical (rod shaped) or even irregular ~\cite{brown2007influence}. 
\begin{figure}[H]
\begin{tikzpicture}[scale=1]
    \draw (-4, 0) node[inner sep=0] {\includegraphics[width=3.3in, trim=0cm 0cm 0cm 0cm, clip]{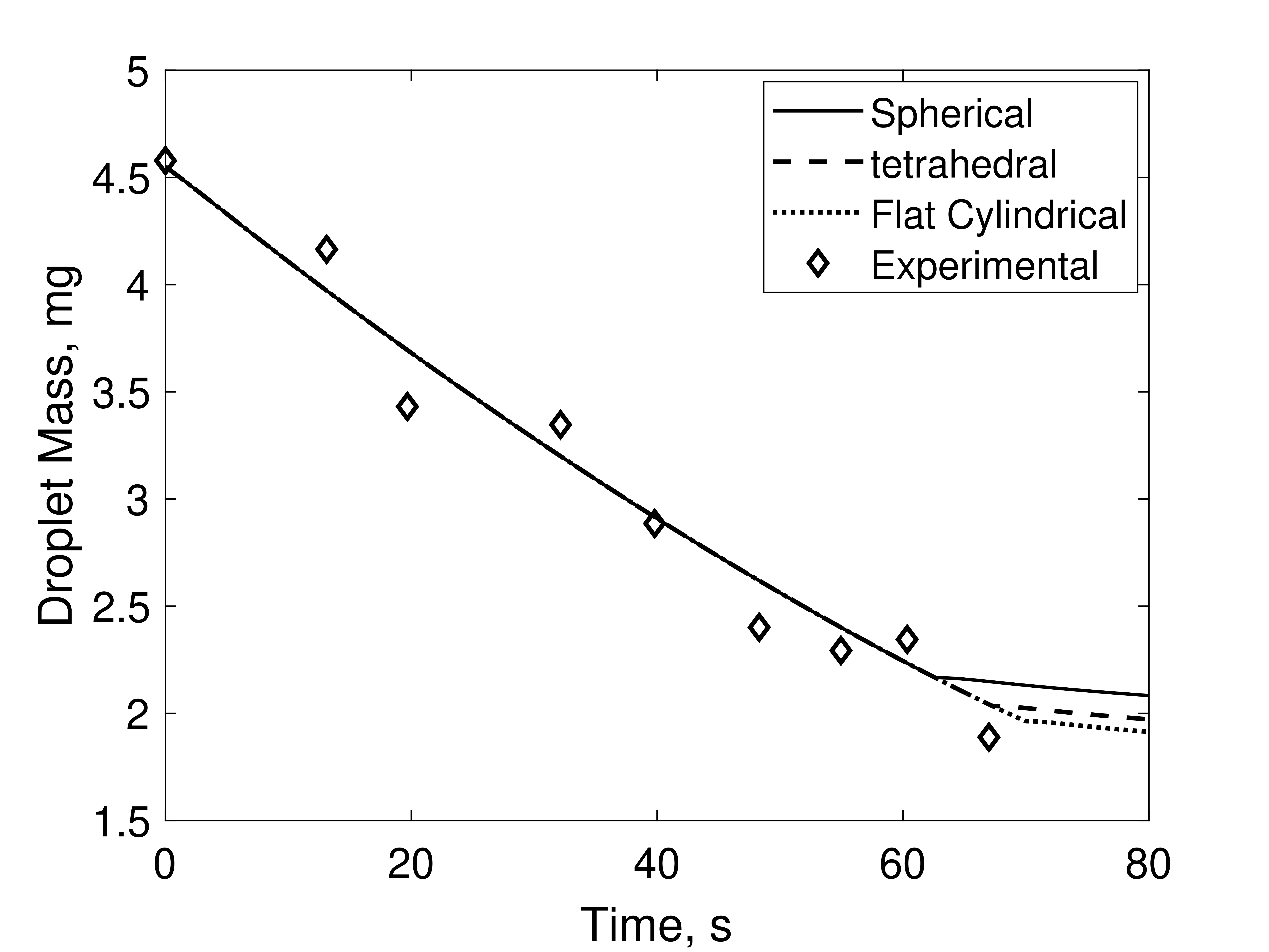}};
    \draw (4, 0) node[inner sep=0] {\includegraphics[width=3.3in, trim=0cm 0cm 0cm 0cm, clip]{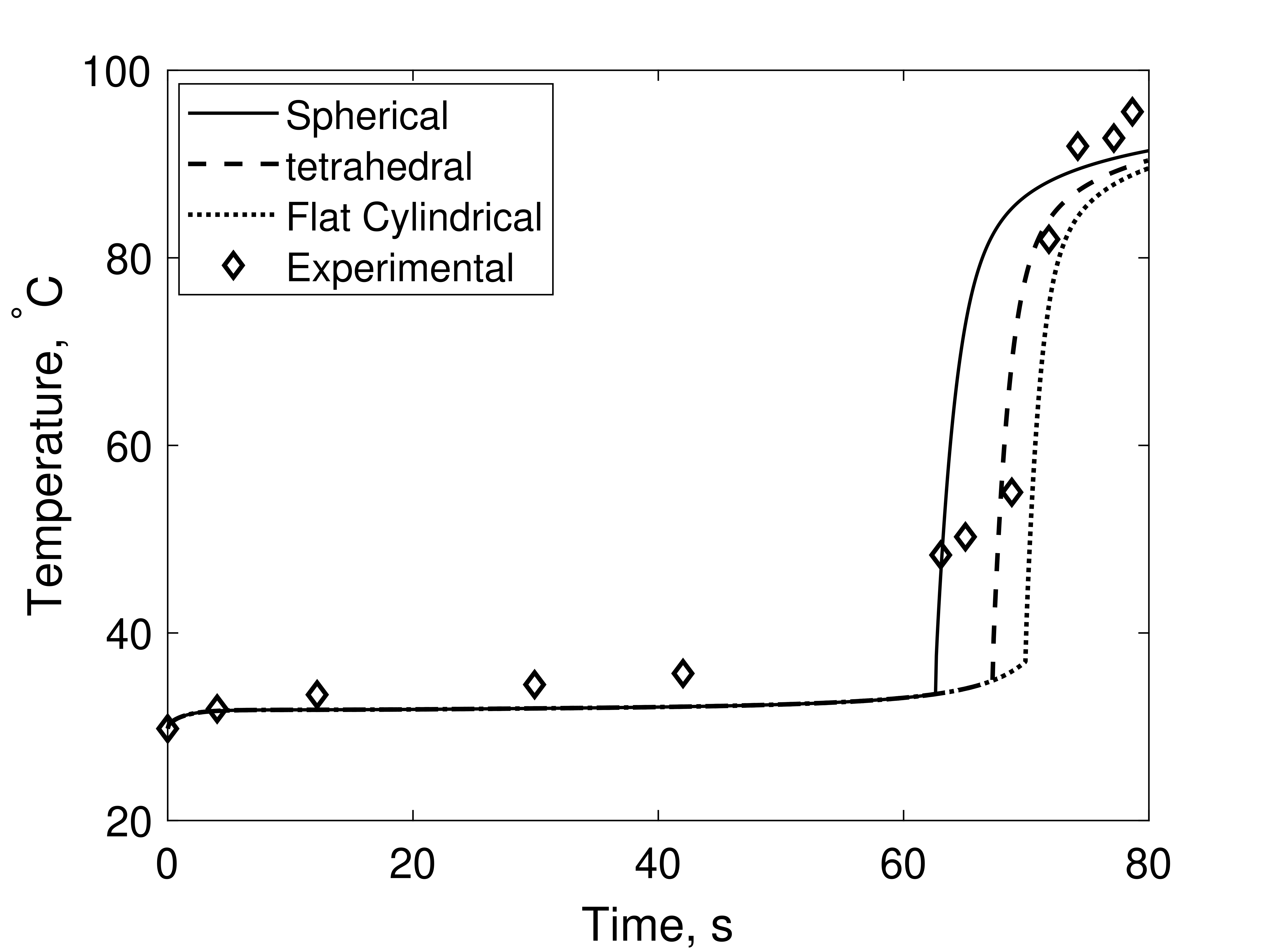}};
    \draw (-7, 3.2) node[scale=1]{$(a)$};
    \draw (1, 3.2) node[scale=1]{$(b)$};
\end{tikzpicture}
\caption{Effect of solid component shape for colloidal silica droplet drying experiment of Ne{\v{s}}i{\'c} and Vodnik~\cite{nevsic1991kinetics} on the evolution of (a)  Mass and (b) Temperature.}
\label{fig:shape}
\end{figure}
When the outermost layer of the droplet is composed of a few layers of particles, the hexagonal close-packing theory~\cite{weisstein2003cubic} can be used to calculate the maximum packing efficiency. To account for different shapes, we calculate the critical solid volume fraction, which is the ratio of the volume occupied by the solid particles on the droplet surface to the total volume. This is essentially the same as the packing efficiency of the solid particles, which is the ratio of the maximum volume that can be occupied by the solid particles to the total volume. While spherical particles have a maximum packing efficiency of 74\%~\cite{hales1998overview}, tetrahedral particles can achieve a packing efficiency of 85\%, and the octahedron shape has a packing efficiency of 95\%~\cite{betke2000densest}. Similarly, solids with very flat cylindrical shapes have a packing efficiency of 91\%~\cite{weisstein2003cubic}. As the packing efficiency increases, the number of solid particles on the surface also increases. This leads to an increase in the critical solid saturation concentration due to the presence of more solid particles on the surface. Figure~\ref{fig:shape} illustrates the impact of various particle shapes on the evolution of mass and temperature. The influence on mass evolution is minimal, with the shape exhibiting a higher critical saturation showing the least moisture removal. However, notable differences are observed in temperature evolution. The transition from the first stage to the second stage appears to be more gradual for the shape with a higher saturation concentration.

\subsubsection{Effect of Crust Pore Size and Discussion on Non-Continuum Effects}

\begin{figure}[H]
\begin{tikzpicture}[scale=1]
    \draw (-4, 0) node[inner sep=0] {\includegraphics[width=3.3in, trim=0cm 0cm 0cm 0cm, clip]{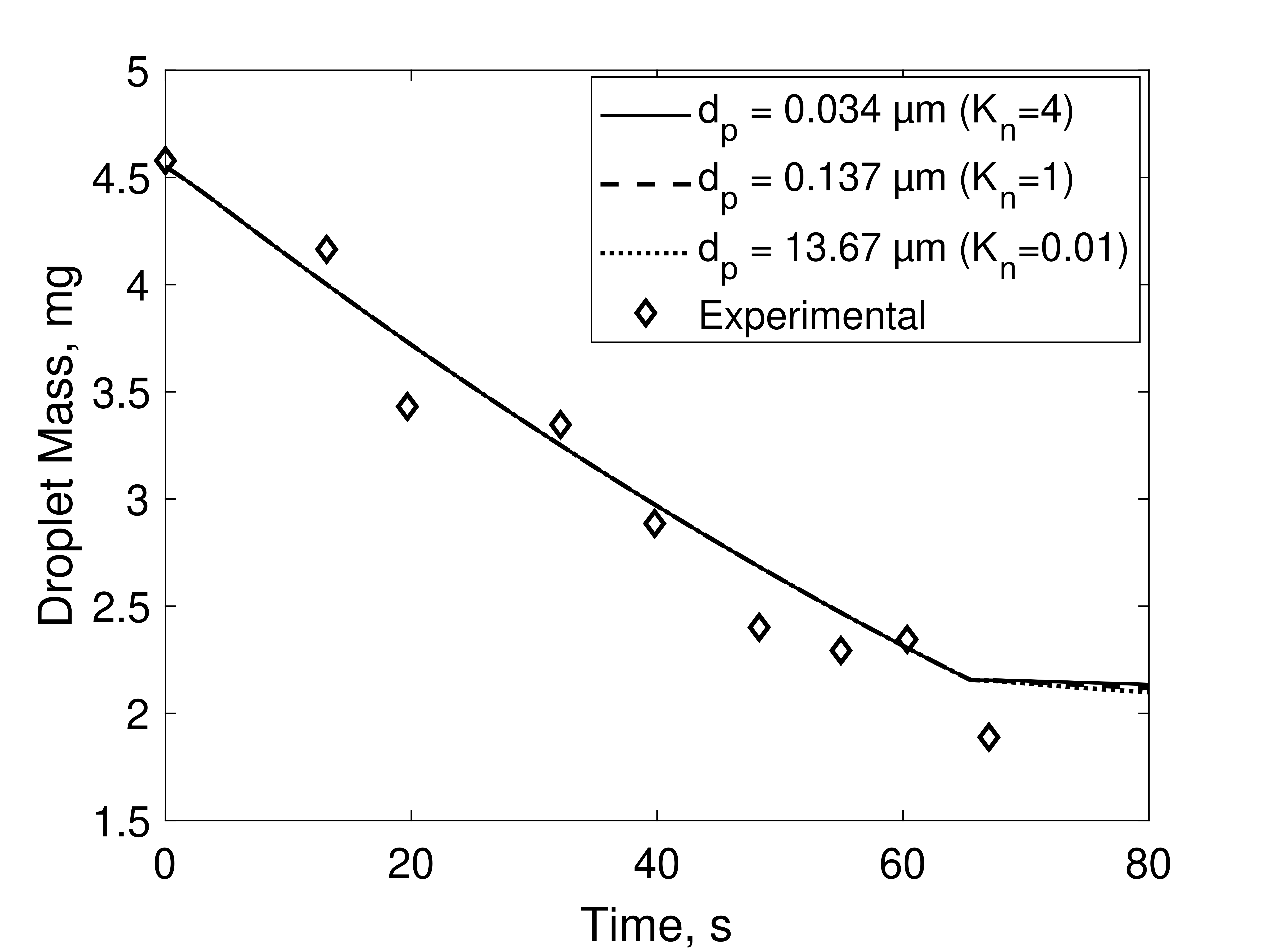}};
    \draw (4, 0) node[inner sep=0] {\includegraphics[width=3.3in, trim=0cm 0cm 0cm 0cm, clip]{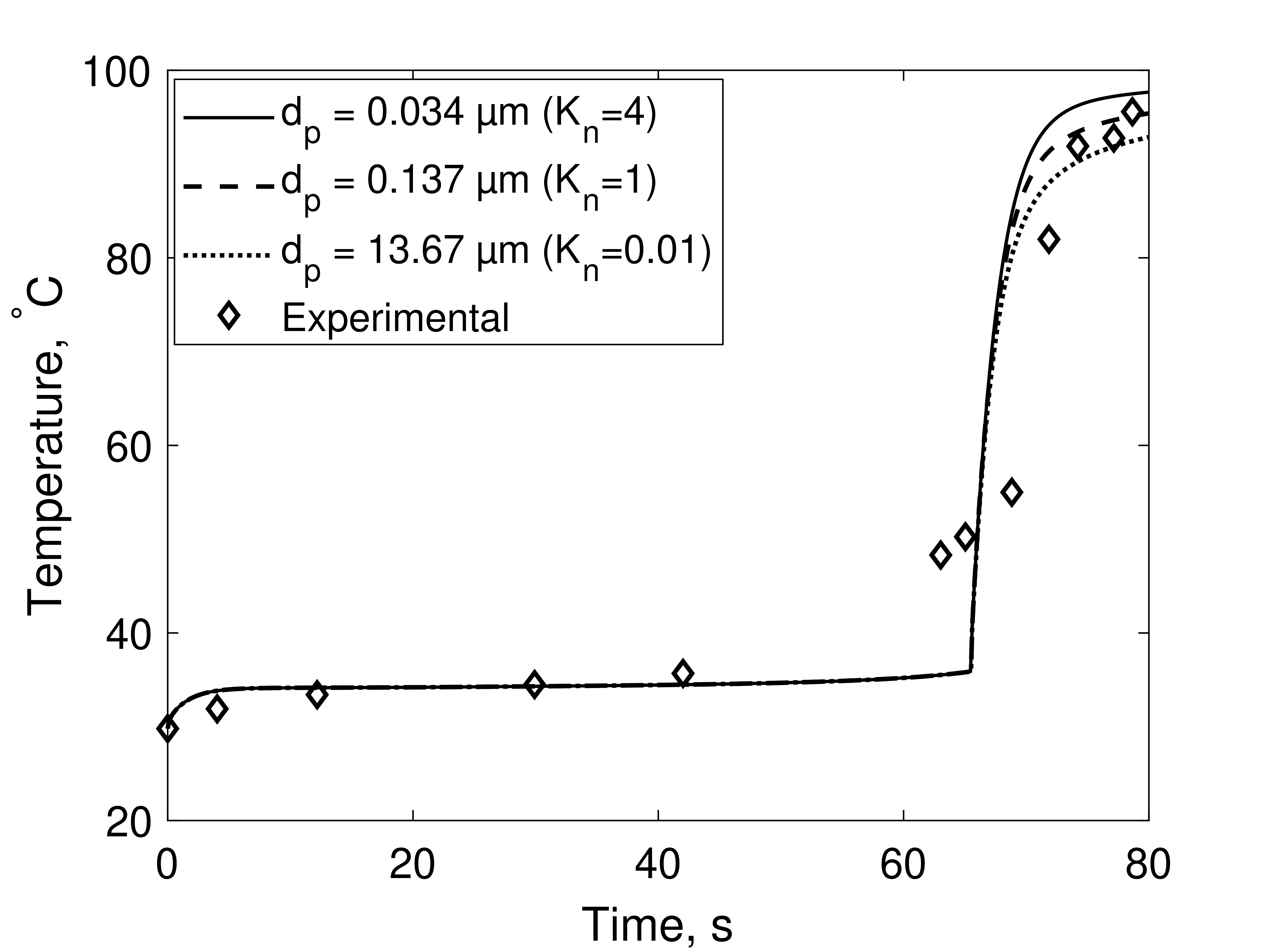}};
    \draw (-7, 3.2) node[scale=1]{$(a)$};
    \draw (1, 3.2) node[scale=1]{$(b)$};
\end{tikzpicture}
\caption{Effect of crust pore size on the evolution of (a)  Mass and (b) Temperature.}
\label{fig:knudsen}
\end{figure}

 The size of the droplets used in spray drying can range anywhere between 1~$mm$ to 20~$\mu m$. With droplet sizes this small, the nature of vapor flow through the crust becomes dependent on the size of the crust pores, $d_p$ and mean free path of the vapor molecules, $\lambda$.
For very small Knudsen number ($Kn = \frac{\lambda}{d_p}$), the pores are much larger compared to the mean free path of the vapor molecules, such that the flow is primarily driven by Fick's diffusion.  However, for larger Knudsen numbers, vapor transport through the pores predominantly includes molecular collisions with walls of the pores. In such a case, the continuum approximations are not valid and a diffusive Knudsen number is used to account for the non-continnum effects. The mean free path of the vapor molecules is calculated using~\cite{nagy2018basic}:

\begin{equation}
    \lambda = \frac{1}{\sqrt{2}}\frac{K_B T}{\pi d_c^2 \overline{p}},
\end{equation}
where $K_B$, $d_c$ and $\overline{p}$ represent the Boltzmann constant, collision diameter of the vapor molecule and mean pressure inside the pores respectively. The collision diameter for the water vapor molecule is assumed to be 0.264 $nm$~\cite{nagy2018basic}. With these values, the initial mean free path of water vapor molecules is calculated to be 0.137~$\mu m$. When non-continuum effects are present, the effective diffusion coefficient is the sum of Fick's diffusion and Knudsen diffusion and is given by~\cite{nagy2018basic}:
\begin{equation}
    D_{eff} = \frac{1}{\frac{1}{D_v} + \frac{1}{D_k}}.
\end{equation}
The Knudsen diffusion coefficient is calculated using~\cite{nagy2018basic}-
\begin{equation}
    D_k = \frac{d_p}{3}\overline{v},
\end{equation}
where $\overline{v}$ is the mean molecular speed given by:
\begin{equation}
    \overline{v} = \sqrt{\frac{8RT}{\pi M}},
\end{equation}
where $R$, $T$ and $M$ are the gas constant, the crust temperature and the molecular weight of the vapor species respectively. We exercise our model for three different pore sizes corresponding to $Kn=$ 0.01, 1 and 4. 
It is interesting to note that while there are negligible differences in the mass transfer profile in Figure~\ref{fig:knudsen}, the temperature evolution for ($K_n = 1$) matches better with the experimental data suggesting that the pore size in the crust could be equal to or smaller than the mean free path of the water vapor molecules. Further, the inclusion of diffusive Knudsen number is shown to considerably affect the temperature profile.

\subsubsection{Effect Of Crust Porosity}
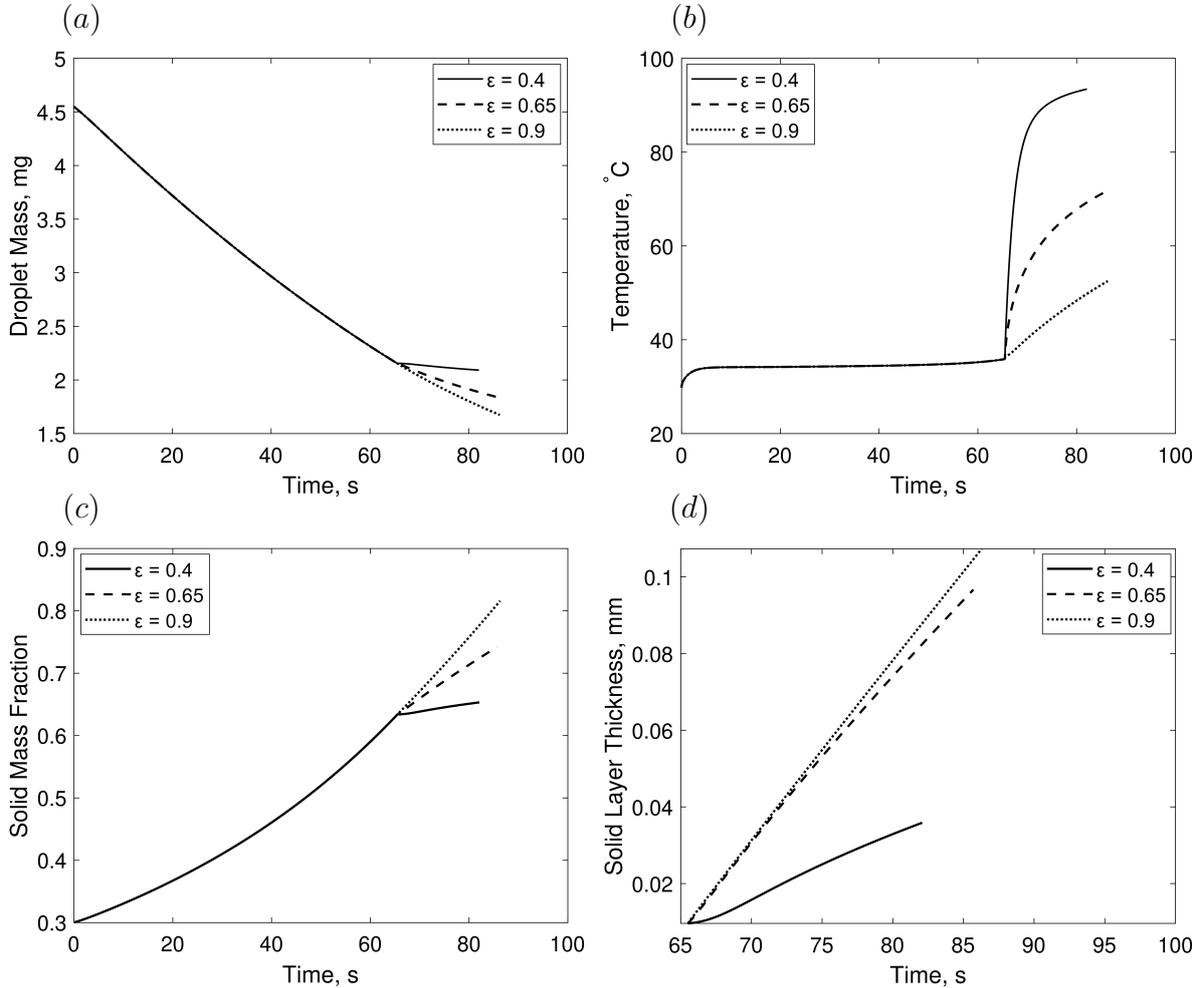
\begin{figure}[H]
\begin{tikzpicture}[scale=1]
    \draw (-4, 0) node[inner sep=0] {\includegraphics[width=3.3in, trim=0cm 0cm 0cm 0cm, clip]{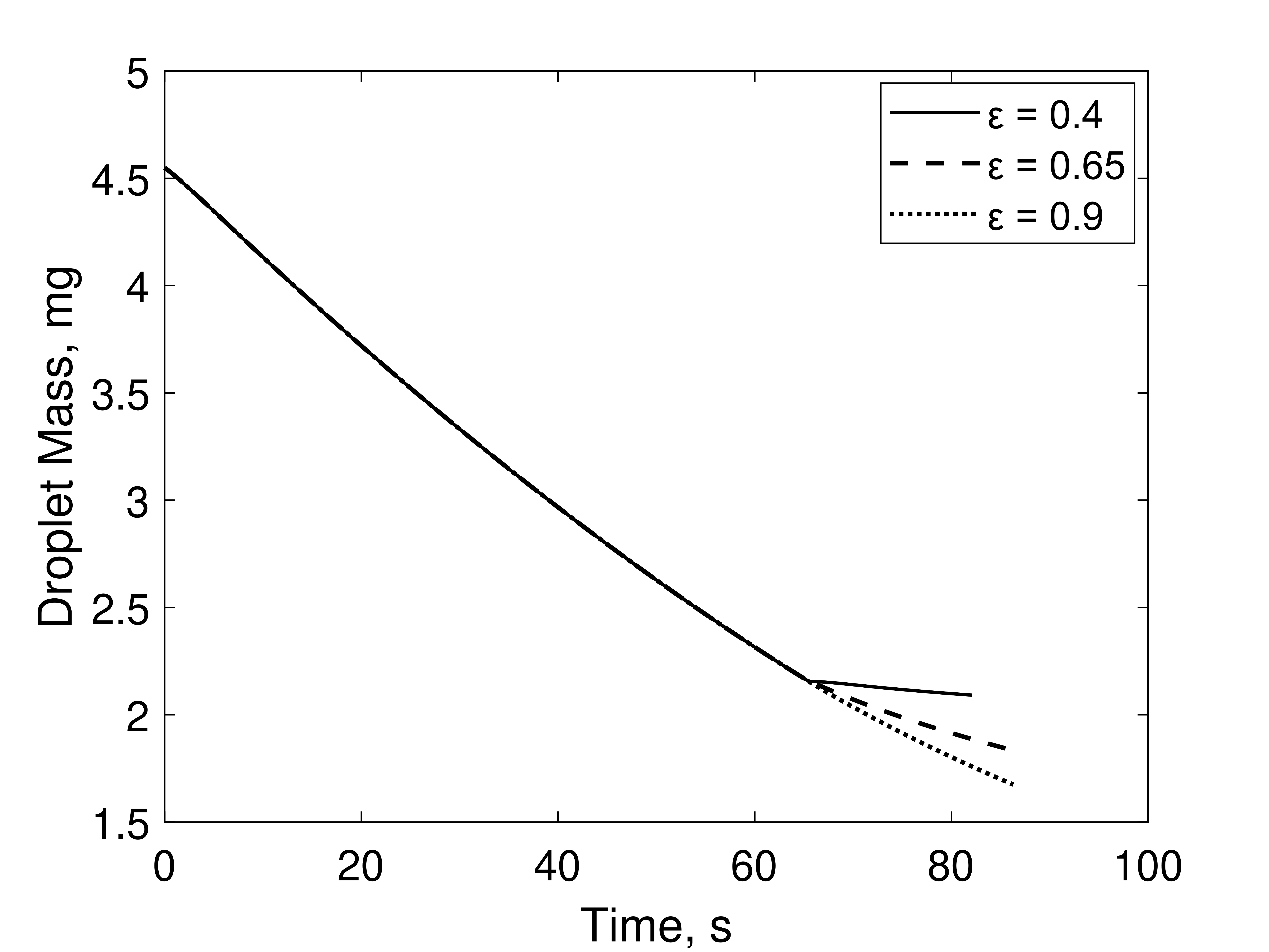}};
    \draw (4, 0) node[inner sep=0] {\includegraphics[width=3.3in, trim=0cm 0cm 0cm 0cm, clip]{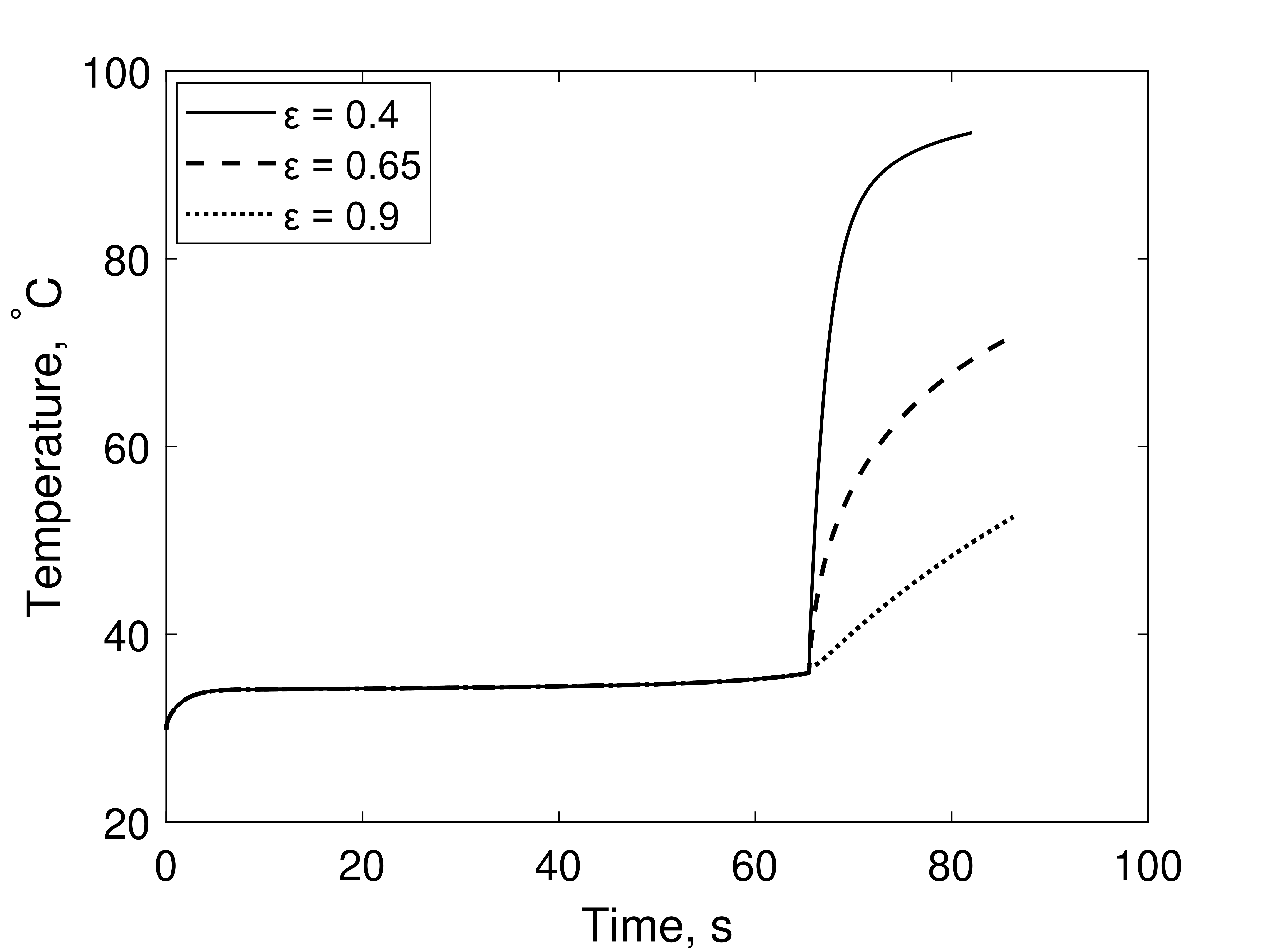}};
    \draw (-4, -6.5) node[inner sep=0] {\includegraphics[width=3.3in, trim=0cm 0cm 0cm 0cm, clip]{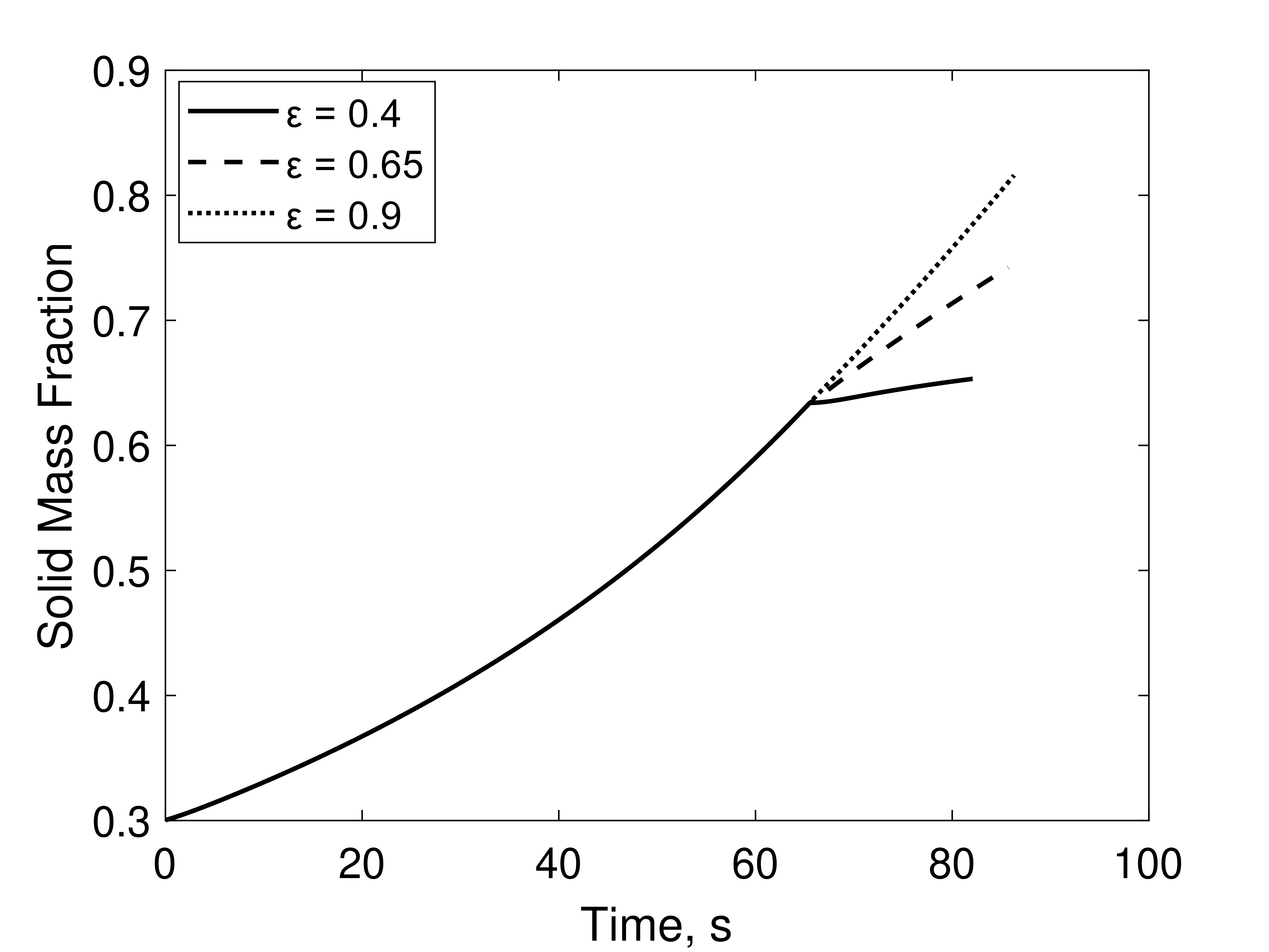}};
    \draw (4, -6.5) node[inner sep=0] {\includegraphics[width=3.3in, trim=0cm 0cm 0cm 0cm, clip]{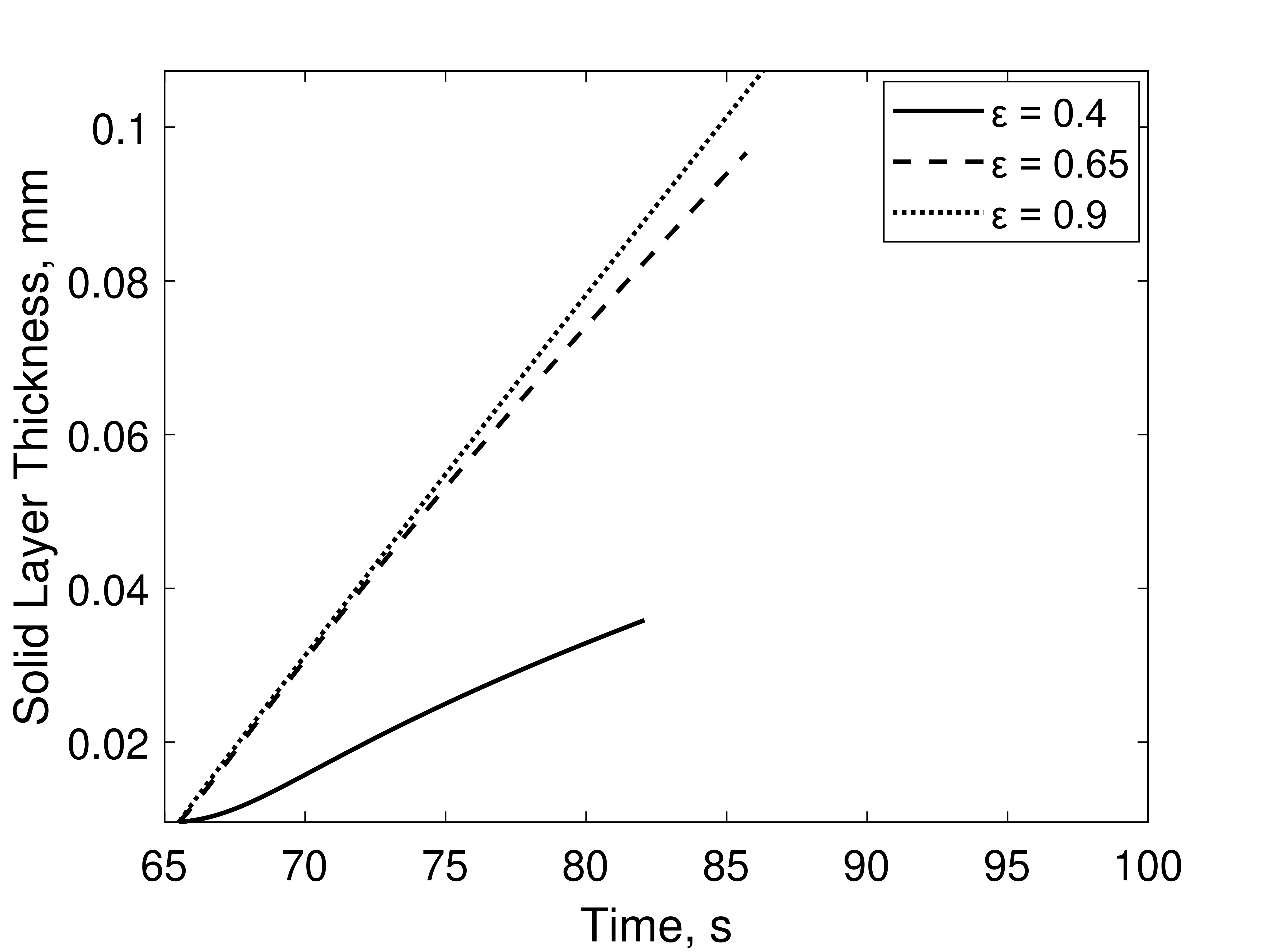}};    
    \draw (-7, 3.2) node[scale=1]{$(a)$};
    \draw (1, 3.2) node[scale=1]{$(b)$};
    \draw (-7, -3.3) node[scale=1]{$(c)$};
    \draw (1, -3.3) node[scale=1]{$(d)$};    
\end{tikzpicture}
\caption{Effect of crust porosity on the evolution of (a)  Mass (b) Temperature (c) Solid mass fraction and (d) Solid layer thickness.}
\label{fig:porosity}
\end{figure}

The porosity of the crust is another factor that affects the transport of vapor through the crust. The porosity of the crust can vary even for the same pore size, depending on the distribution of the pores. To understand the role of porosity in the drying process, we excercise our model with different porosity values. The results, shown in Figure~\ref{fig:porosity}(a), indicate that a highly porous crust leads to a higher rate of moisture removal. This is expected because a porous crust allows more vapor to pass through, while a less porous crust restricts the flow of vapor, resulting in a lower rate of moisture removal. Figure~\ref{fig:porosity}(b) demonstrates that the temperature is inversely related to the porosity of the crust. A more porous crust has a lower solid content, which reduces the overall thermal conductivity of the crust and consequently lowers the temperature. The porosity of the crust also affects the solid mass fraction and the growth of the interface radius, as shown in Figure~\ref{fig:porosity}(c) and (d). A less porous crust hinders the transfer of vapor from the wet-core, leading to slower crust growth and, consequently, a lower solid mass fraction and crust thickness.

\vspace{-0.1in}

\subsubsection{Spatial Variation of Temperature within the Droplet and Discussion on Biot Number}\label{sec:biot}
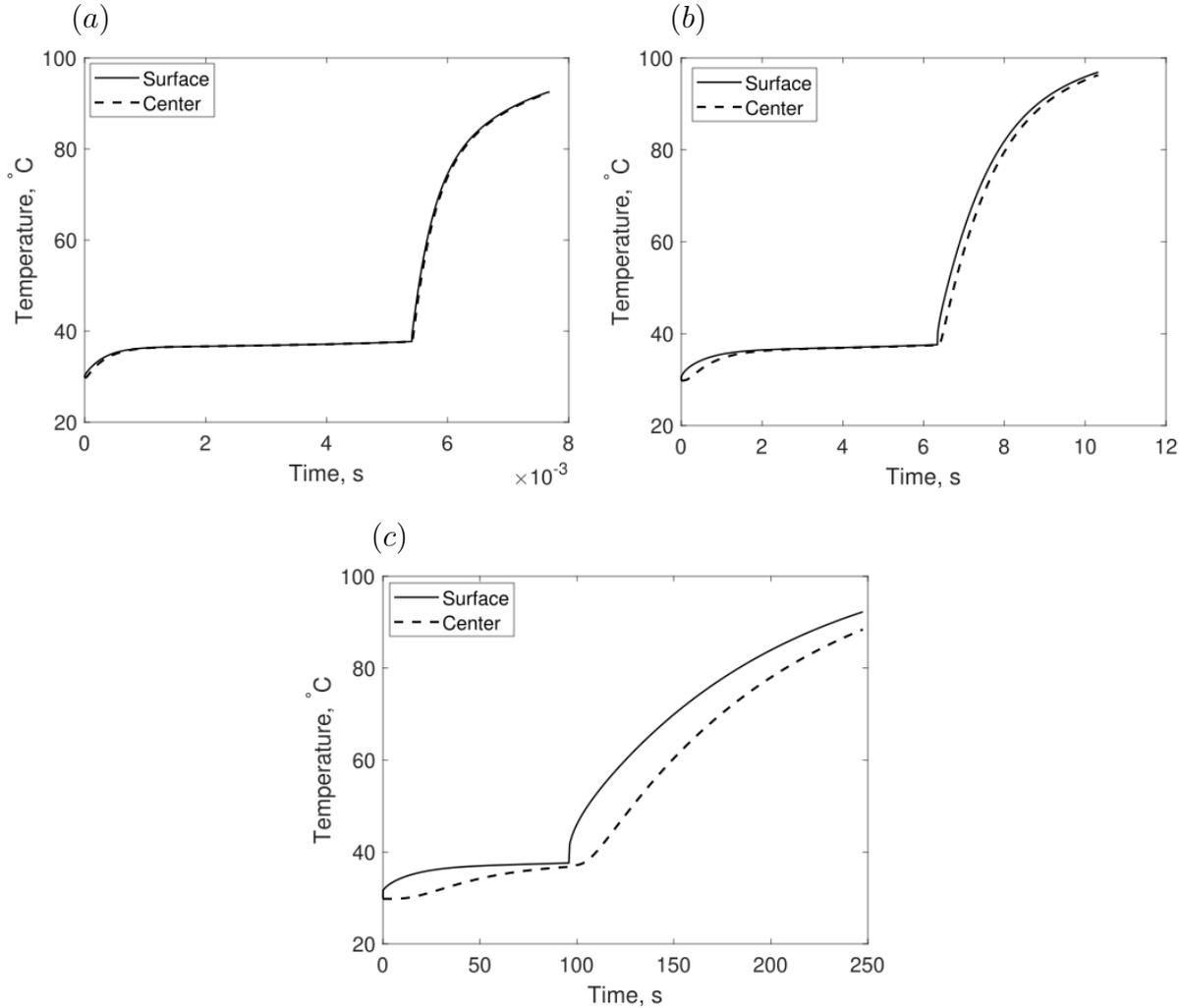
\begin{figure}[H]
\begin{tikzpicture}[scale=1]
    \draw (-4, 0) node[inner sep=0] {\includegraphics[width=3.3in, trim=0cm 0cm 0cm 0cm, clip]{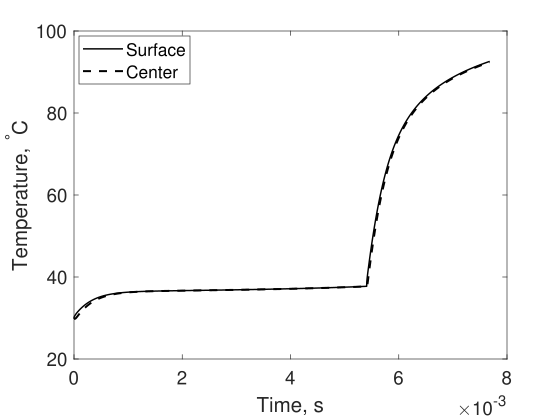}};
    \draw (4, 0) node[inner sep=0] {\includegraphics[width=3.3in, trim=0cm 0cm 0cm 0cm, clip]{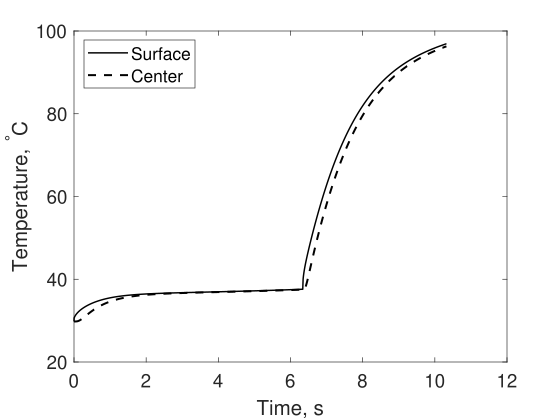}};
    \draw (0, -7) node[inner sep=0] {\includegraphics[width=3.3in, trim=0cm 0cm 0cm 0cm, clip]{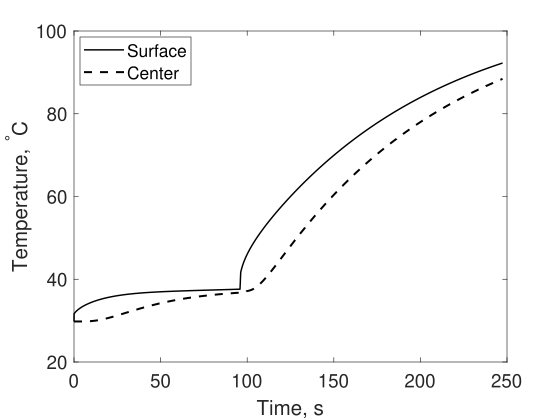}};    
    \draw (-7, 3.2) node[scale=1]{$(a)$};
    \draw (1, 3.2) node[scale=1]{$(b)$};
    \draw (-3, -3.8) node[scale=1]{$(c)$};     
\end{tikzpicture}
\caption{Evolution of temperature at the droplet surface and center for droplets with three different Biot numbers. $Bi=$ (a)  0.06  (b) 0.2 (c) 0.5.}
\label{fig:biot}
\end{figure}

The size of the droplet has a significant impact on the temperature distribution within it. For small droplets with a Biot number ($ Bi = \frac{h R_d}{k_d}$) less than 0.1, it is reasonable to assume a uniform temperature distribution. However, for larger droplets that do not meet this criterion, it is necessary to consider the temperature distribution profile inside the droplet. Previous studies have supported the assumption of uniform temperature distribution during drying~\cite{parti1974mathematical,nevsic1991kinetics,handscomb2009new,jayanthi1993modeling}. This simplifies the implementation of the model and is suitable for spray drying conditions with high drying temperature and gas velocity. In such conditions, the droplet will lose moisture and shrink rapidly, resulting in a decrease in the Biot number. However, in operations like air suspension particle coating in dairy applications, which are carried out at lower temperatures, a temperature gradient may exist within the droplet for a longer period of time. Figure~\ref{fig:biot}(a),(b) and (c) illustrate the first and second stages of drying for three colloidal silica droplets with Biot number values of 0.06, 0.2, and 0.5, respectively. The plots clearly show that a high temperature gradient exists between the droplet surface and center at high Biot numbers, while these differences are negligible for Biot numbers less than 0.1. 

\subsection{Effect of Drying Conditions}
The objective of this section is to clarify the impact of various drying conditions on the process of drying. Specifically, we investigate the influence of drying conditions on the formation of the solid crust. Ultimately, we create a regime map that can be used to predict whether the particle will be solid or hollow.

\subsubsection{Effect of Drying Gas Relative Humidity}
\begin{figure}[H]
\begin{tikzpicture}[scale=1]
    \draw (-4, 0) node[inner sep=0] {\includegraphics[width=3.3in, trim=0cm 0cm 0cm 0cm, clip]{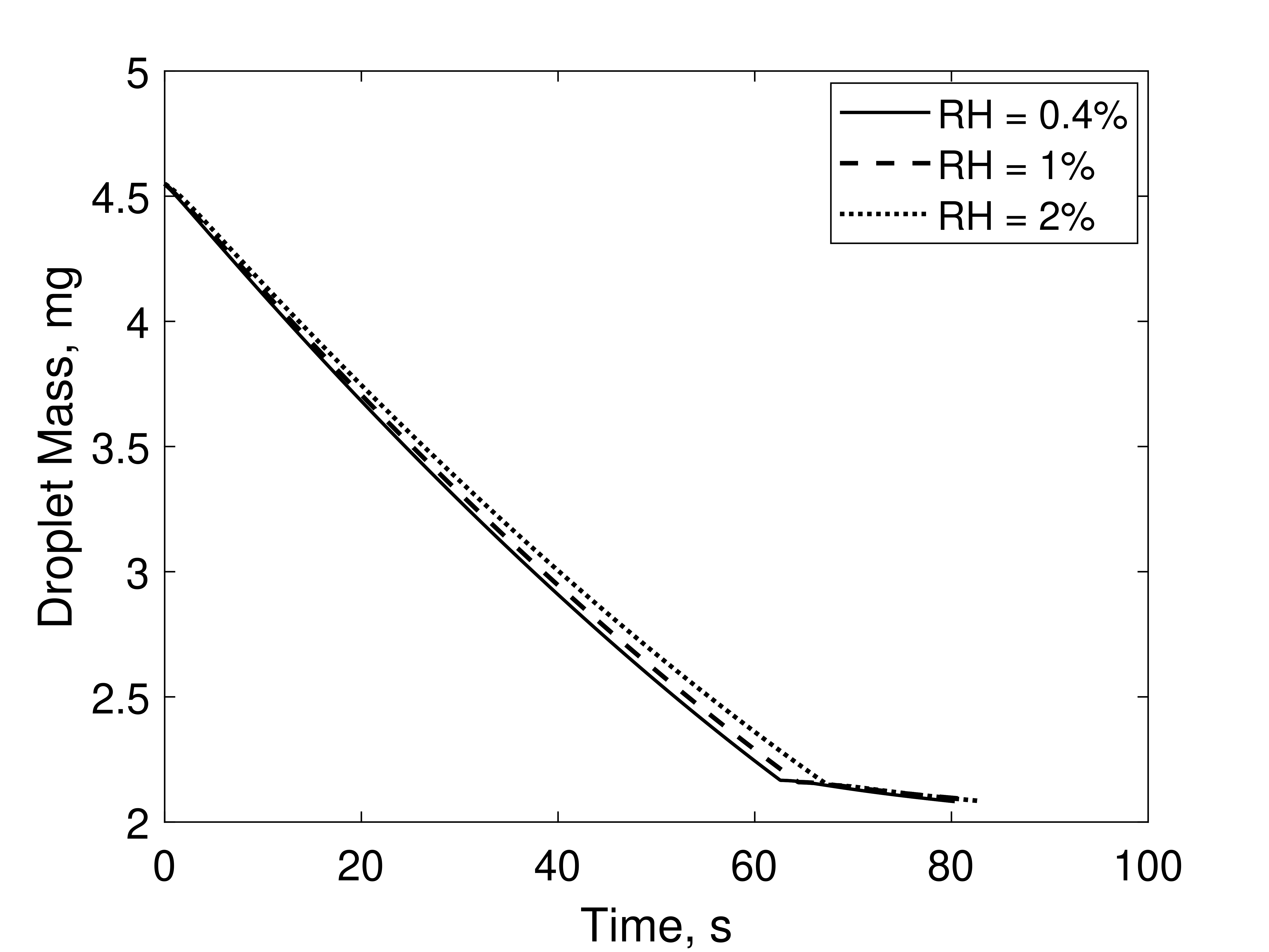}};
    \draw (4, 0) node[inner sep=0] {\includegraphics[width=3.3in, trim=0cm 0cm 0cm 0cm, clip]{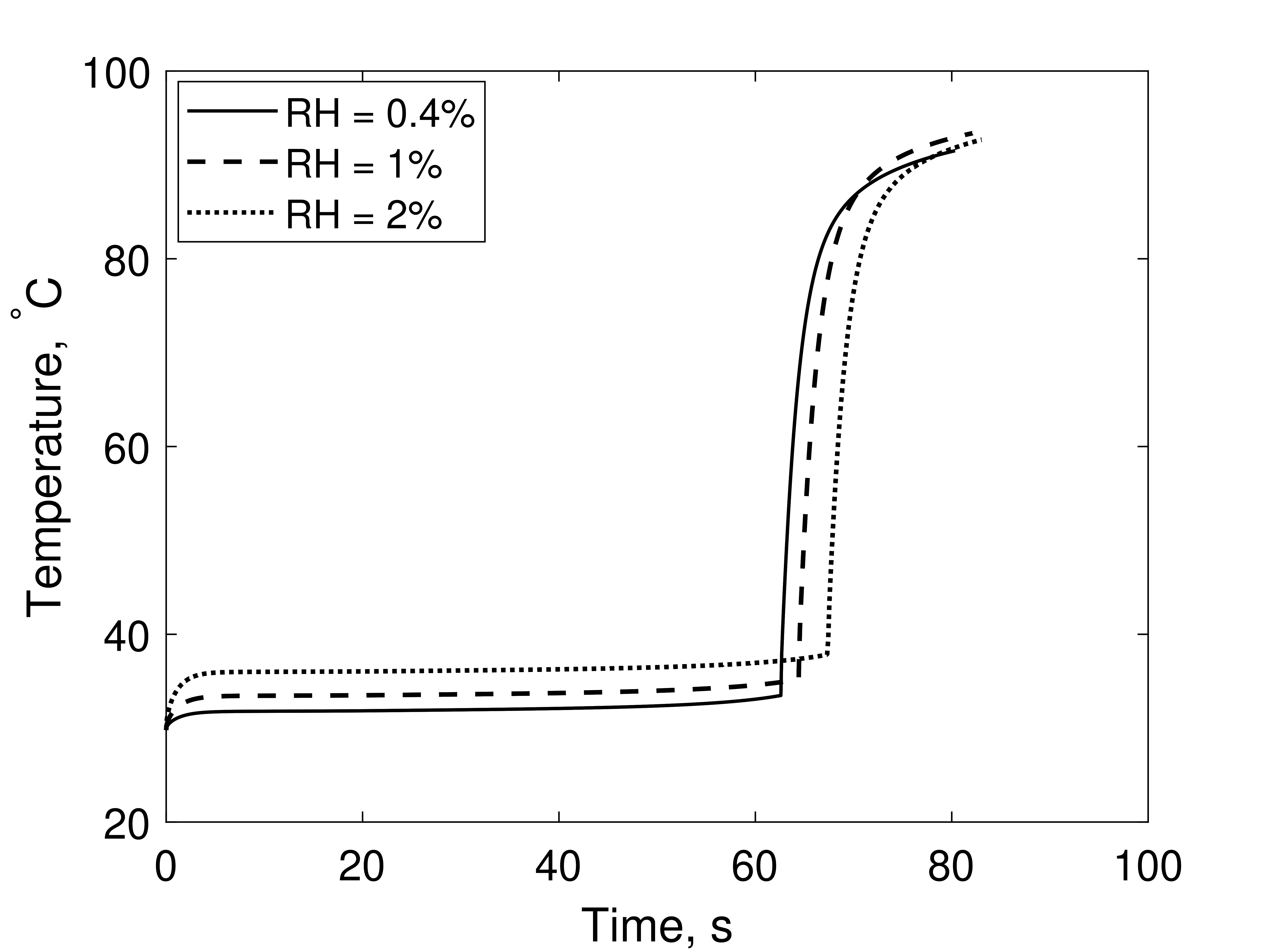}};
    \draw (-4, -6.5) node[inner sep=0] {\includegraphics[width=3.3in, trim=0cm 0cm 0cm 0cm, clip]{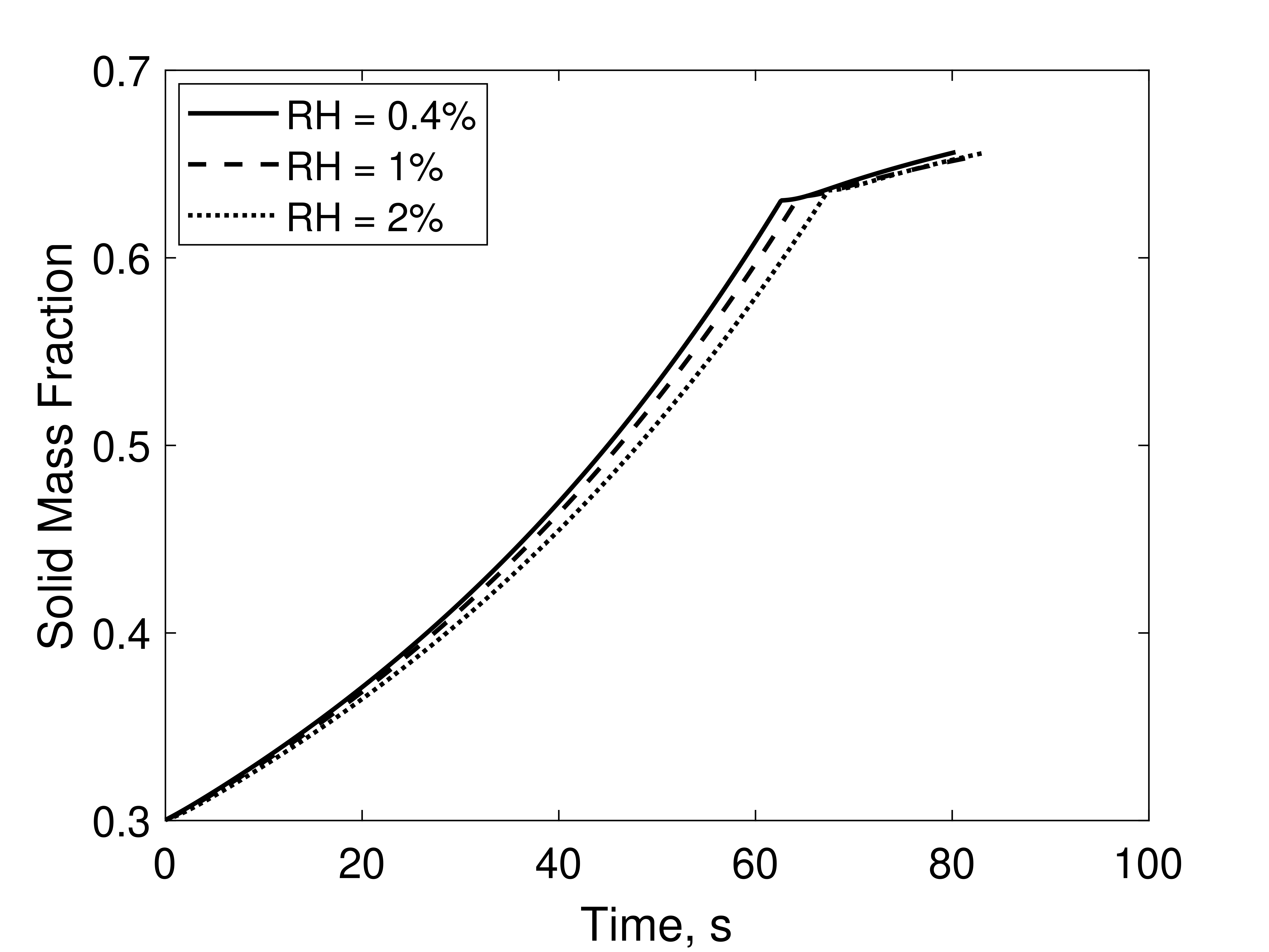}};
    \draw (4, -6.5) node[inner sep=0] {\includegraphics[width=3.3in, trim=0cm 0cm 0cm 0cm, clip]{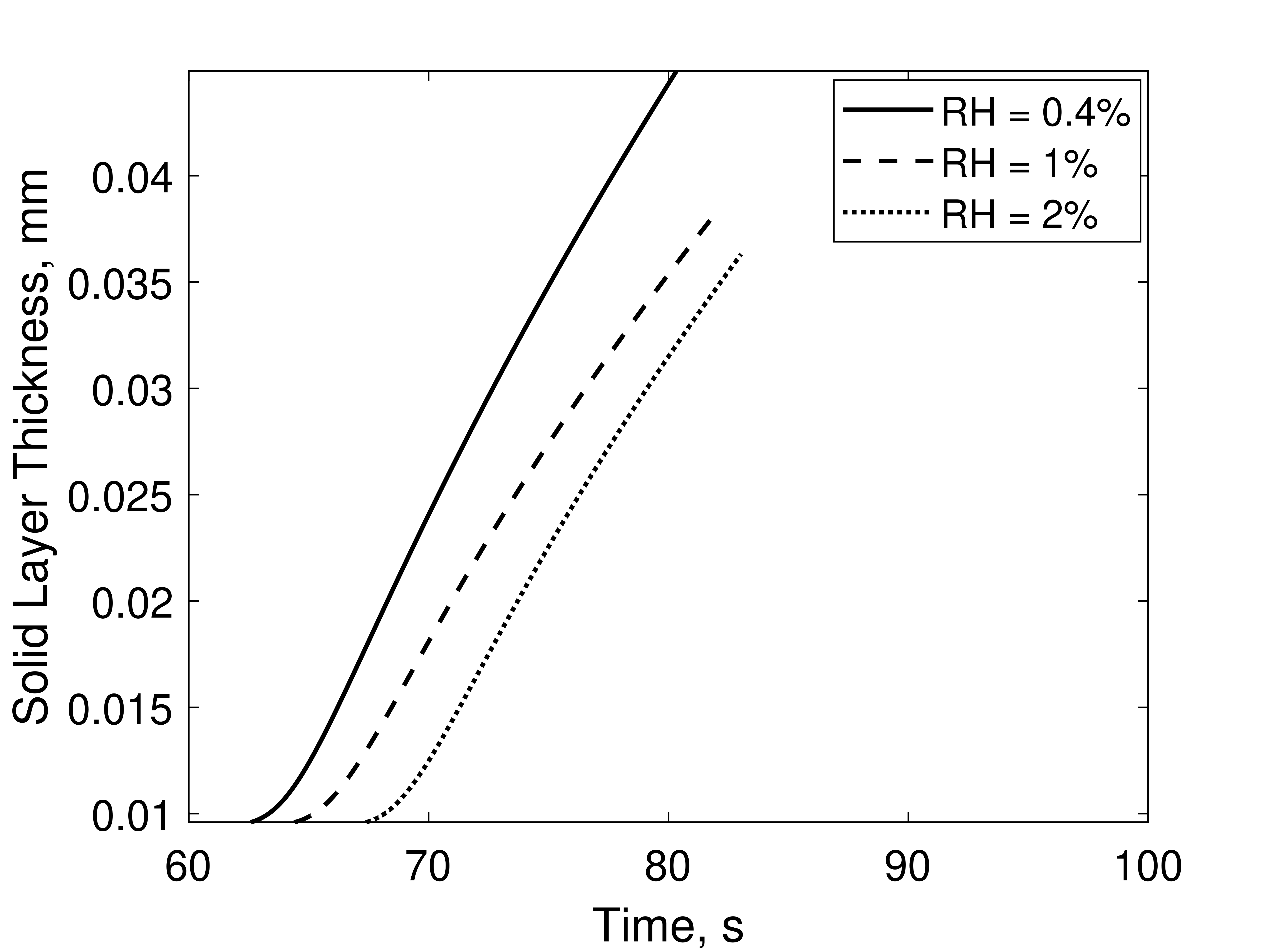}};    
    \draw (-7, 3.2) node[scale=1]{$(a)$};
    \draw (1, 3.2) node[scale=1]{$(b)$};
    \draw (-7, -3.3) node[scale=1]{$(c)$};
    \draw (1, -3.3) node[scale=1]{$(d)$};    
\end{tikzpicture}
\caption{Effect of relative humidity on drying of a  colloidal silica droplet. Figures show the evolution of (a)  Mass (b) Temperature (c) Solid mass fraction (d) Solid layer thickness.}
\label{fig:RH}
\end{figure}
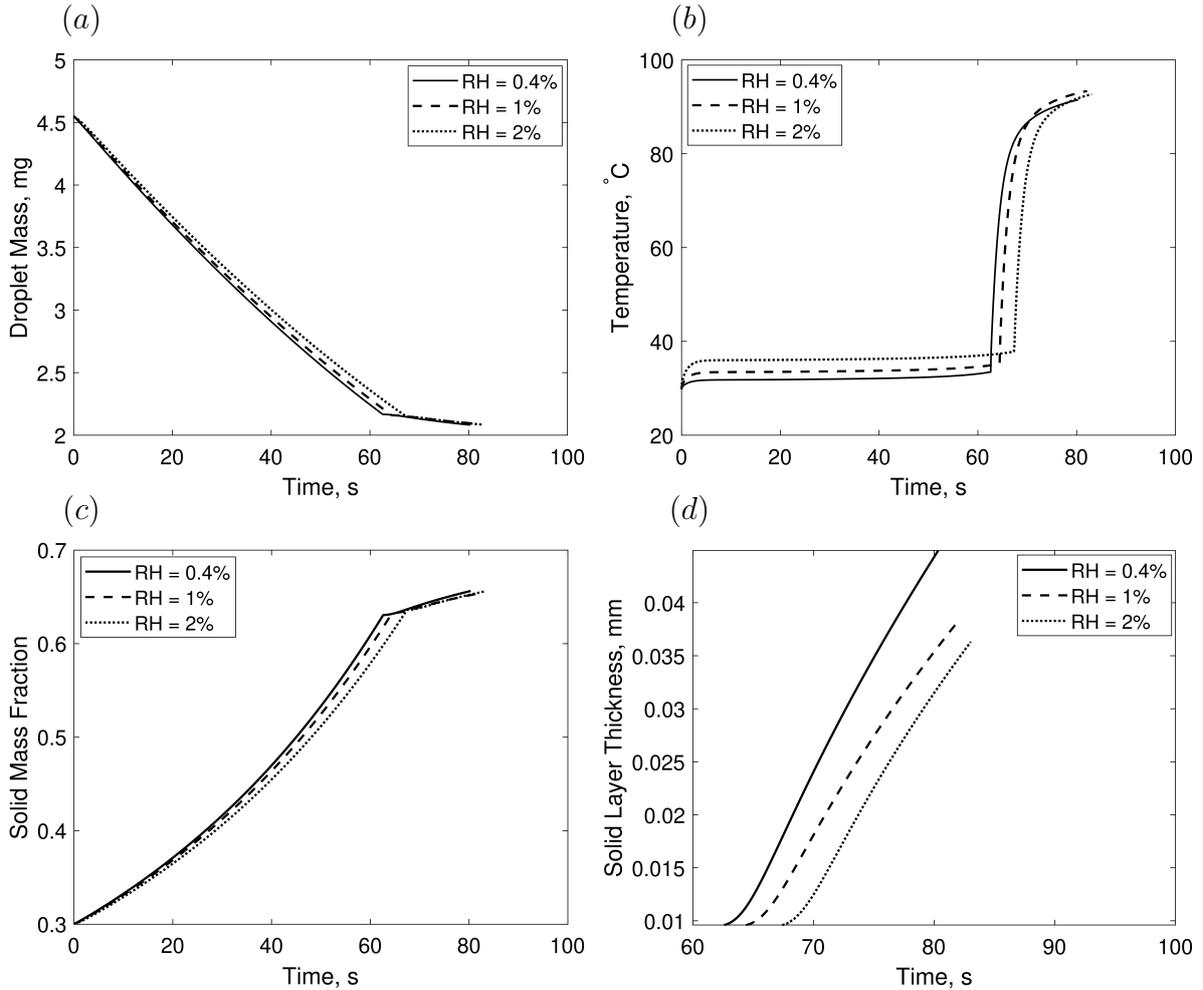

The drying gas's relative humidity can vary significantly within the drying chamber, resulting in different conditions for the droplets depending on their location. For example, the relative humidity near the atomizer nozzle of the chamber is high due to the high density of droplets in that area compared to other regions in the chamber. Therefore, it is crucial to understand the impact of relative humidity on the drying process. Figure~\ref{fig:RH} illustrates how changes in relative humidity affect the temperature and mass of the droplets. In this study, we initially considered a baseline relative humidity value of 0.4\%, as used in the experiments conducted by Ne{\v{s}}i{\'c} and Vodnik~\cite{nevsic1991kinetics}, and then increased it to 1\% and 2\%. It can be observed that relative humidity has only a moderate effect on the temperature and mass evolution of the droplets. A higher relative humidity makes it more difficult to remove moisture from the droplets, resulting in a lower mass transfer rate, as shown in the plot. Additionally, a higher relative humidity leads to a higher wet bulb temperature, which is consistent with the obtained results. Relative humidity has minimal impact on the moisture removal rate during the second stage of droplet drying, as depicted in Figure~\ref{fig:RH}(a). This is also evident in the evolution of the solid mass fraction shown in Figure~\ref{fig:RH}(c), where a higher mass fraction is observed for a relative humidity of 0.4\% during the first stage, but similar progressions are observed during the second stage due to negligible differences in moisture removal rates. The evolution of the solid layer thickness in Figure~\ref{fig:RH}(d) exhibits a similar trend. Although crust formation occurs at different times due to varying transition times from the first stage to the second stage, the overall slope of the plots remains the same, suggesting an equal growth rate of the crust thickness.


\vspace{-0.1in}

\subsubsection{Effect of Drying Gas Temperature}

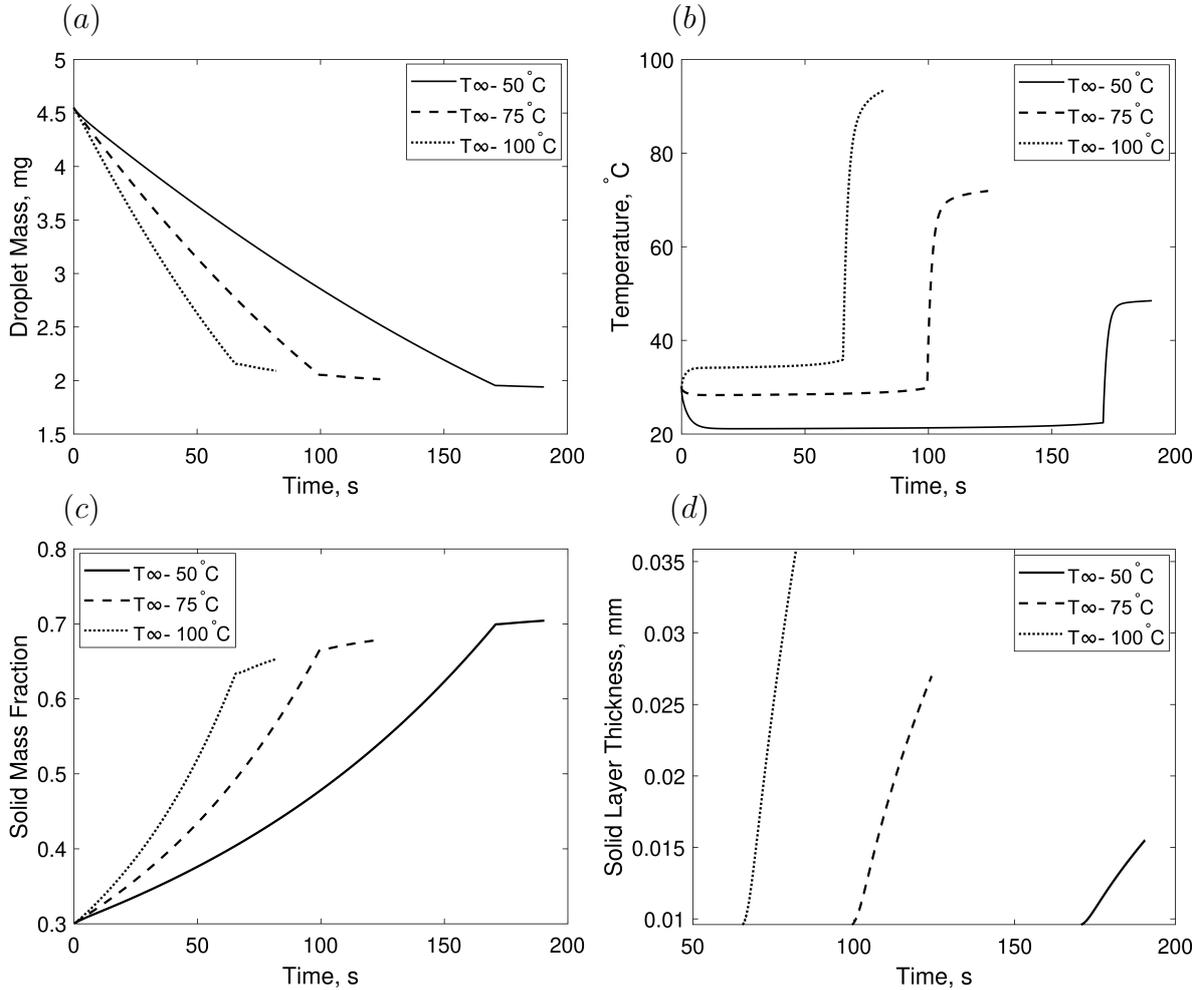
\begin{figure}[H]
\begin{tikzpicture}[scale=1]
    \draw (-4, 0) node[inner sep=0] {\includegraphics[width=3.3in, trim=0cm 0cm 0cm 0cm, clip]{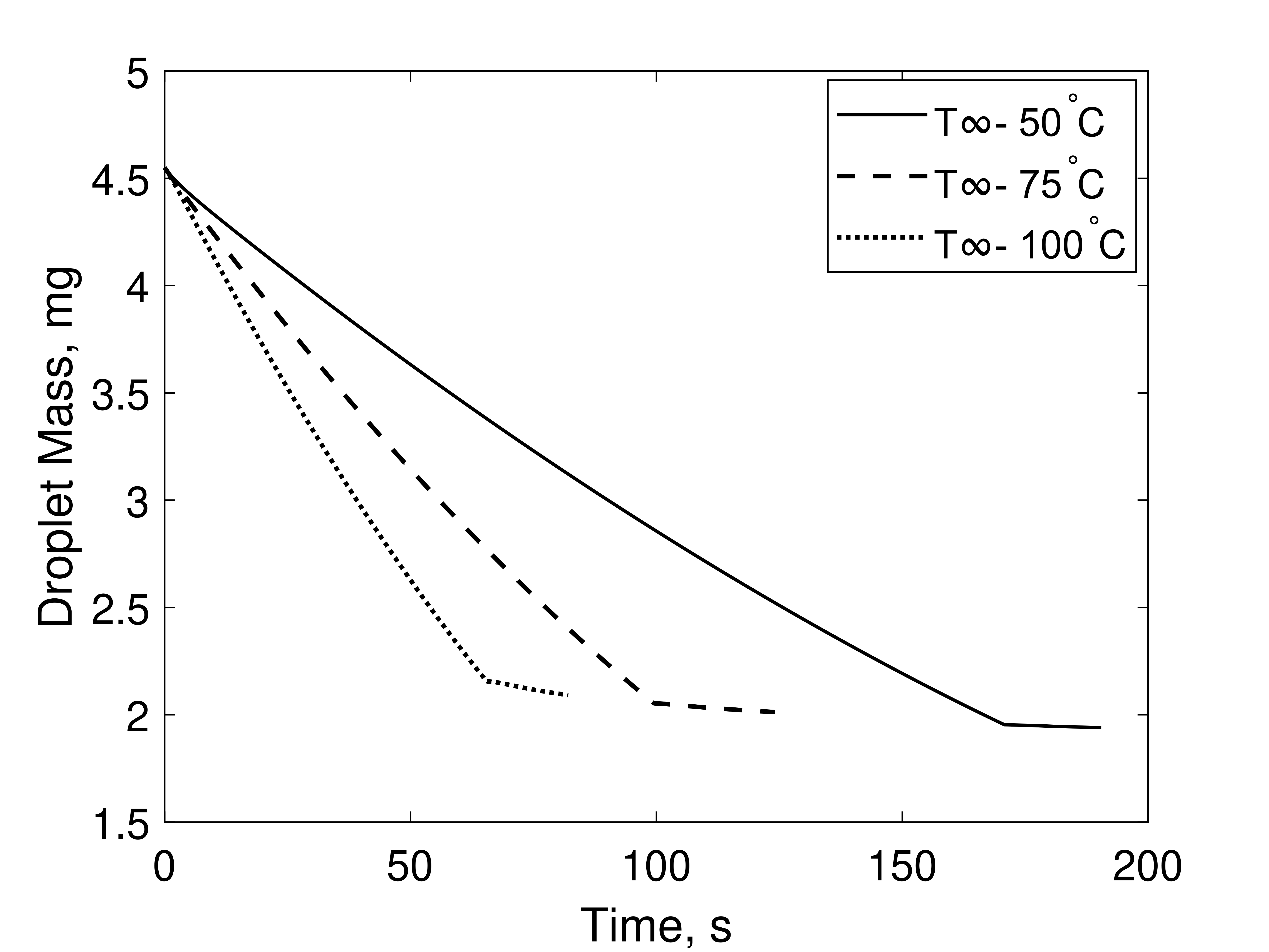}};
    \draw (4, 0) node[inner sep=0] {\includegraphics[width=3.3in, trim=0cm 0cm 0cm 0cm, clip]{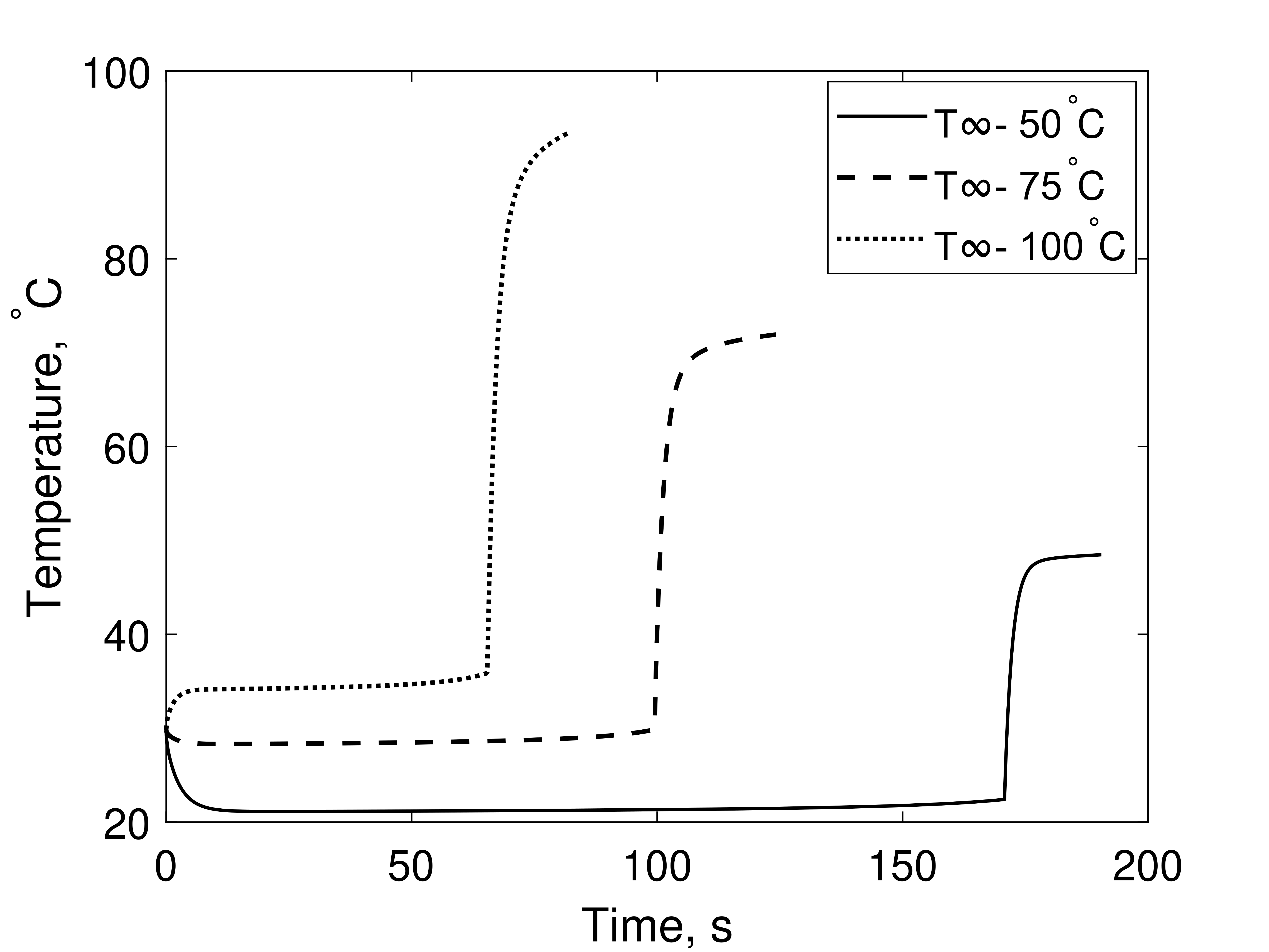}};
    \draw (-4, -6.5) node[inner sep=0] {\includegraphics[width=3.3in, trim=0cm 0cm 0cm 0cm, clip]{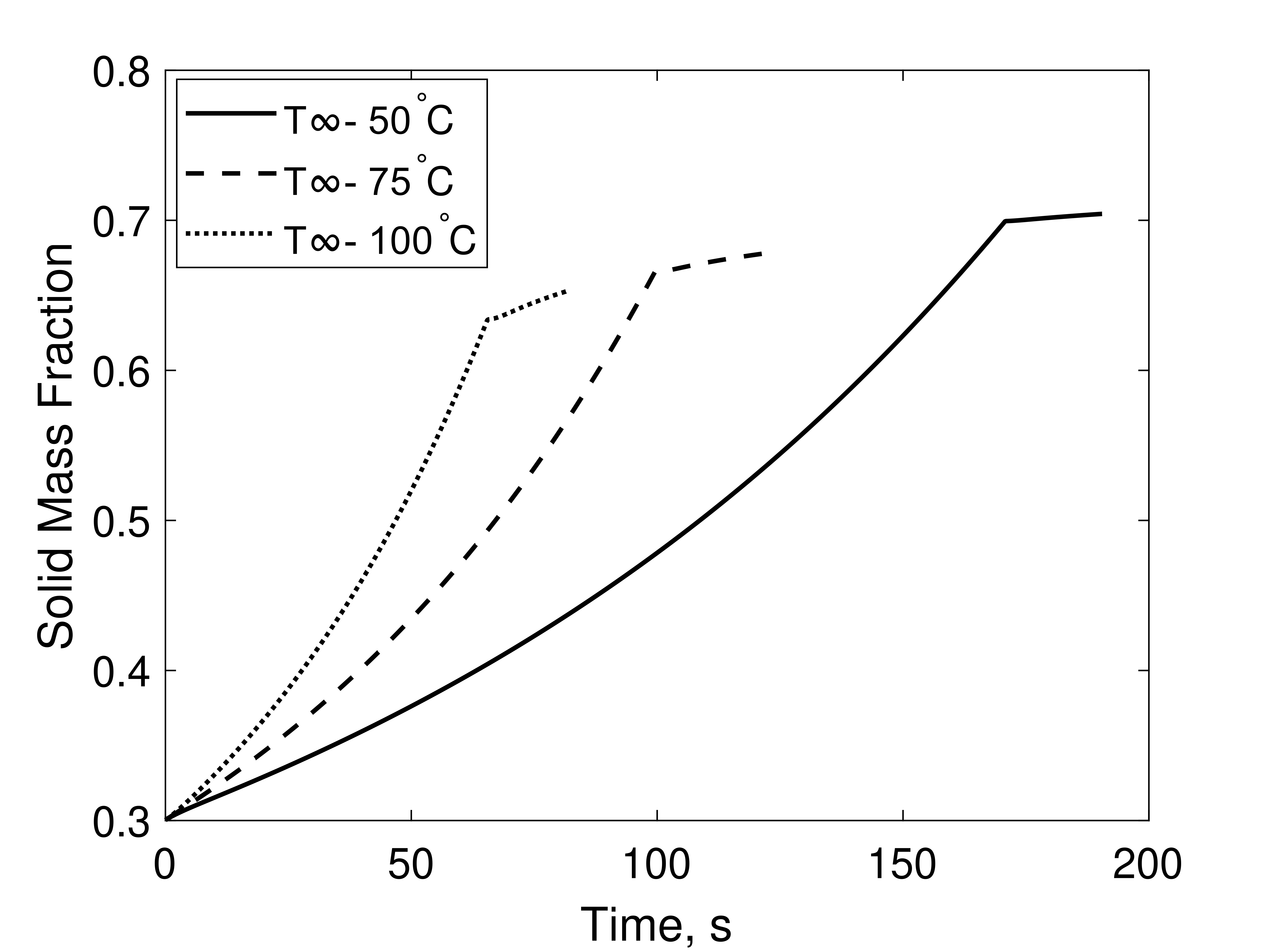}};
    \draw (4, -6.5) node[inner sep=0] {\includegraphics[width=3.3in, trim=0cm 0cm 0cm 0cm, clip]{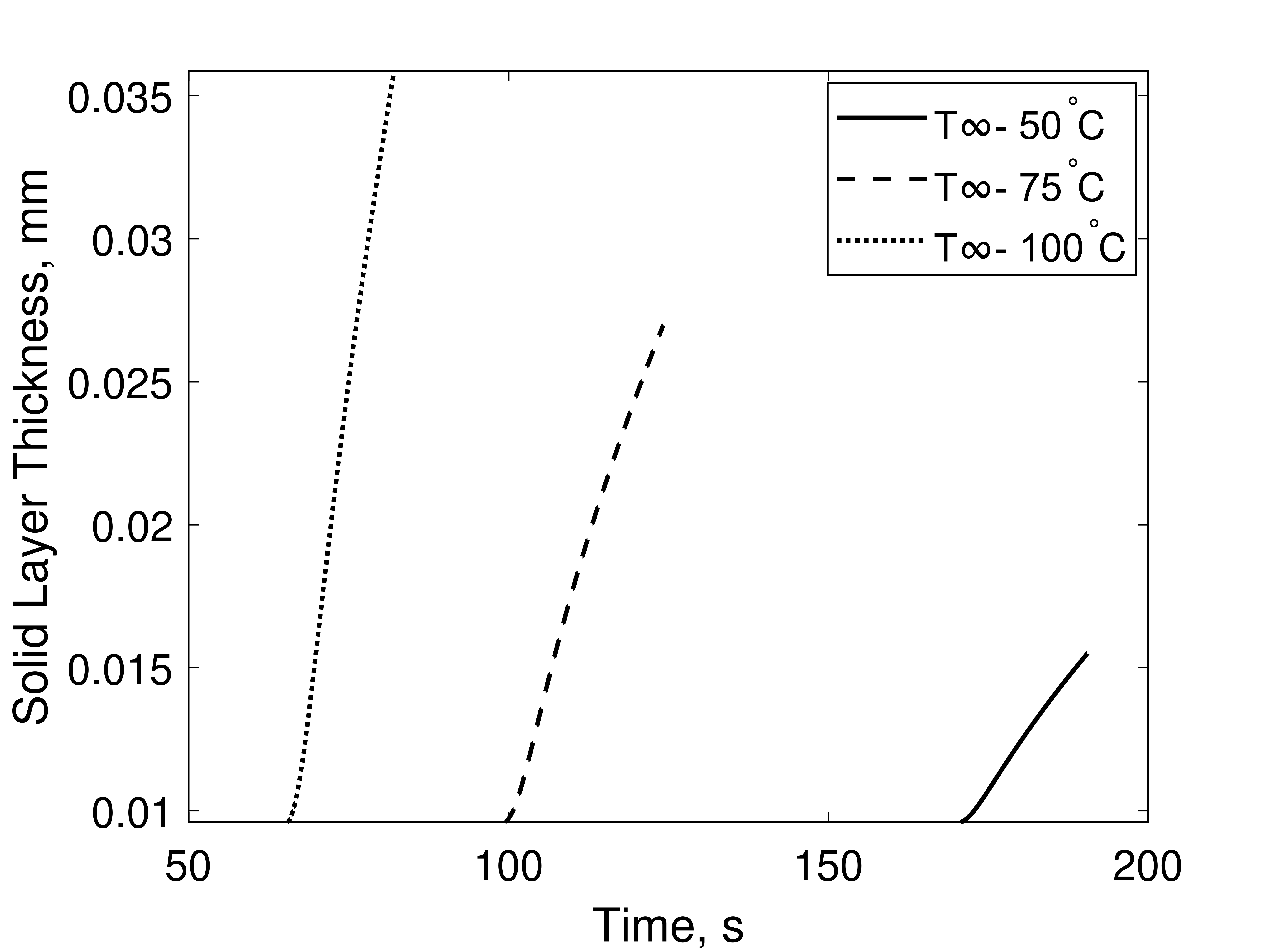}};    
    \draw (-7, 3.2) node[scale=1]{$(a)$};
    \draw (1, 3.2) node[scale=1]{$(b)$};
    \draw (-7, -3.3) node[scale=1]{$(c)$};
    \draw (1, -3.3) node[scale=1]{$(d)$};     
\end{tikzpicture}
\caption{Effect of drying gas temperature on drying of a  colloidal silica droplet. Figures show the evolution of (a) mass, (b) temperature, (c) solid mass fraction and (d) Solid layer thickness.}
\label{fig:temperature}
\end{figure}

The temperature of the gas used for drying is typically a crucial external parameter that can significantly affect the characteristics of the particles after drying. Figure~\ref{fig:temperature} illustrates the impact of three different droplet temperatures (50 $^\circ$C, 75 $^\circ$C, and 100 $^\circ$C) on the evolution of droplet temperature and mass. Various drying temperatures have noticeable effects. It is evident that a higher drying temperature speeds up the drying process by enhancing moisture removal. Moreover, a higher drying temperature raises the wet bulb temperature. Once the second stage of drying begins, the droplet temperature rapidly increases towards the air temperature. This is due to a decrease in the evaporation rate caused by the development of a crust, which introduces additional resistance. A higher ambient temperature would facilitate moisture removal during both stages of droplet drying, resulting in a higher fraction of solid mass at elevated drying temperatures, as shown in Figure~\ref{fig:temperature}(c). Furthermore, a higher temperature would lead to faster crust growth due to increased vapor transfer from the core, as depicted in Figure~\ref{fig:temperature}(d).

\vspace{-0.1in}

\subsubsection{Effect of Drying Gas Velocity}
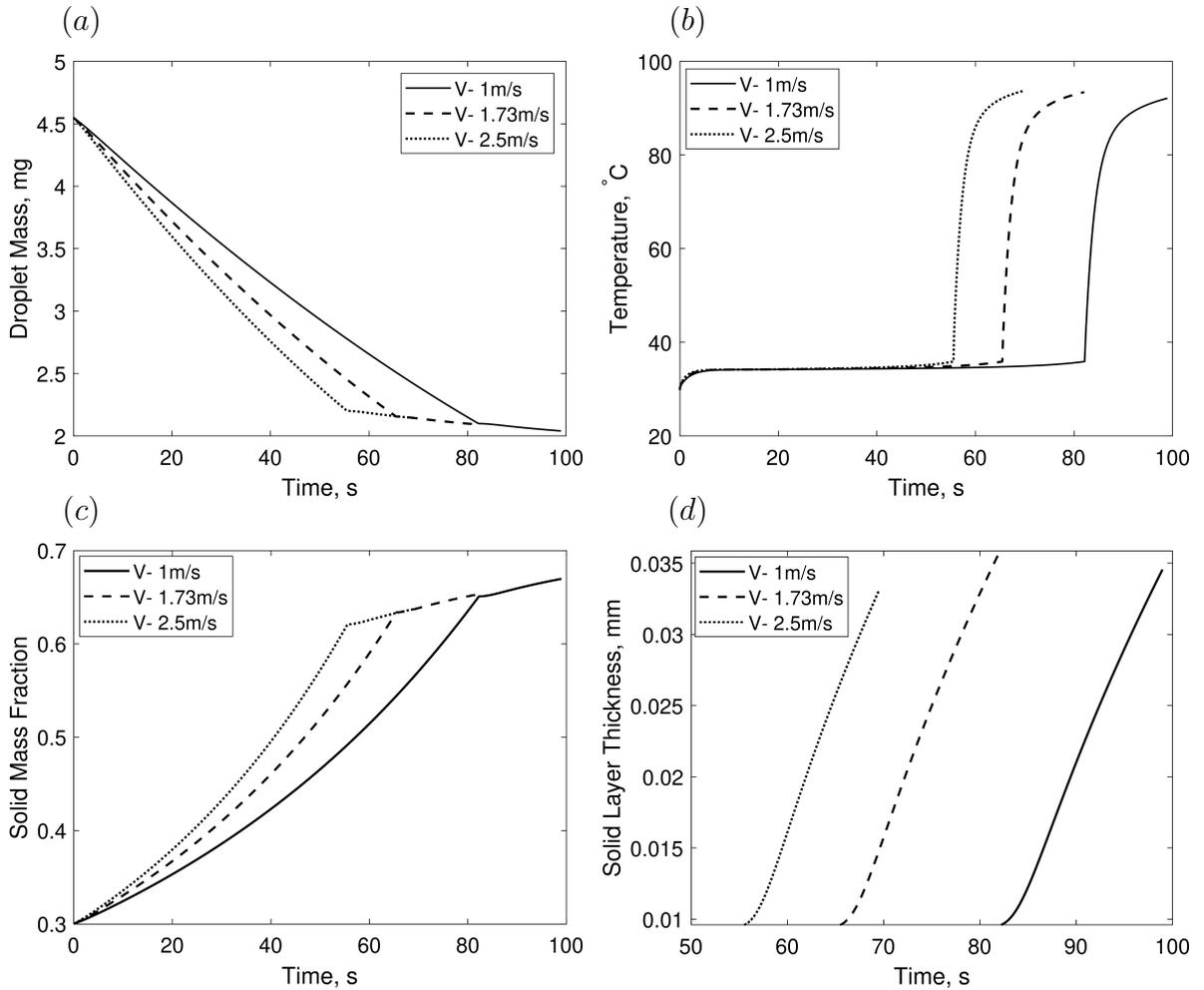
\begin{figure}[H]
\begin{tikzpicture}[scale=1]
    \draw (-4, 0) node[inner sep=0] {\includegraphics[width=3.3in, trim=0cm 0cm 0cm 0cm, clip]{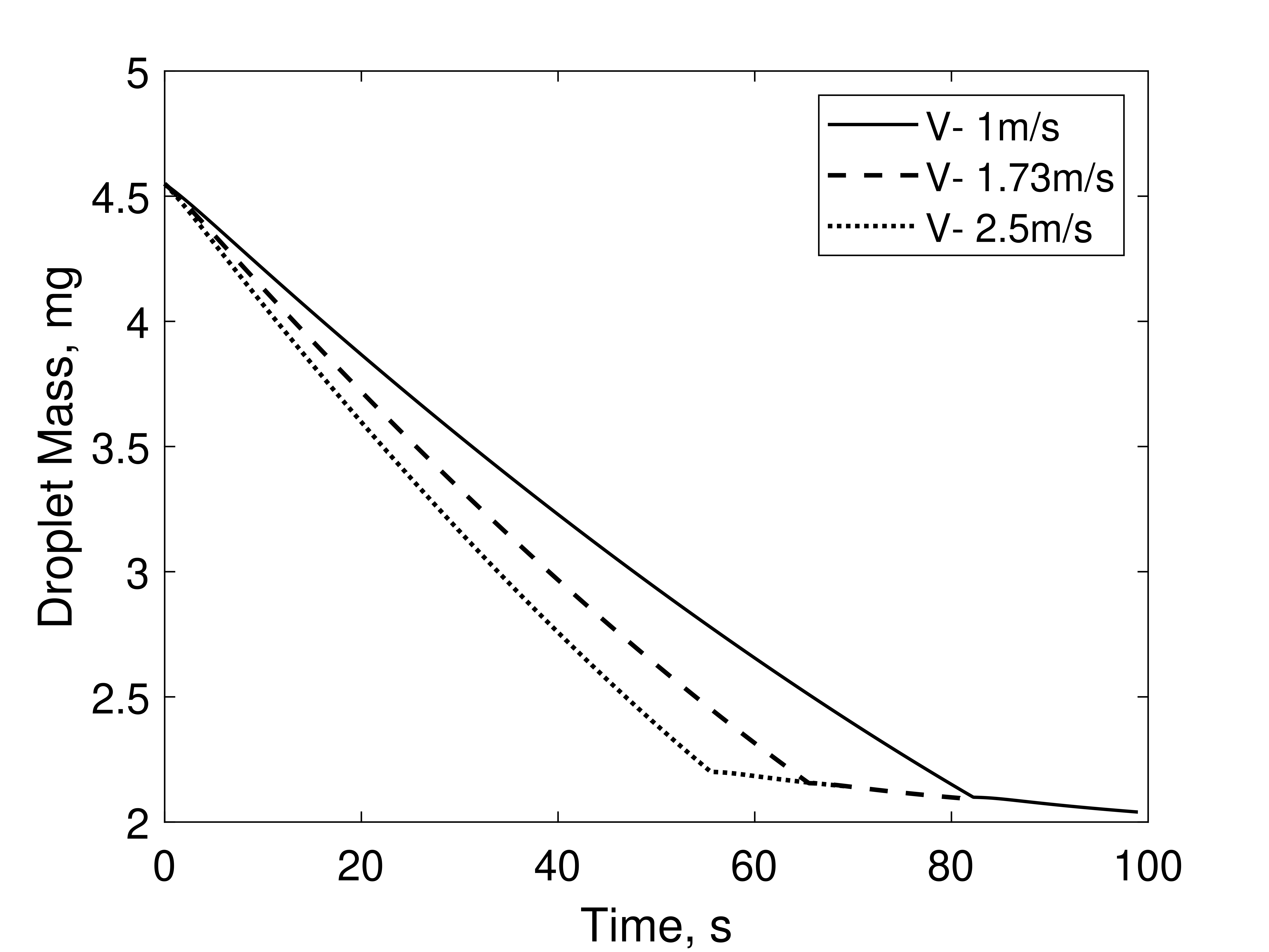}};
    \draw (4, 0) node[inner sep=0] {\includegraphics[width=3.3in, trim=0cm 0cm 0cm 0cm, clip]{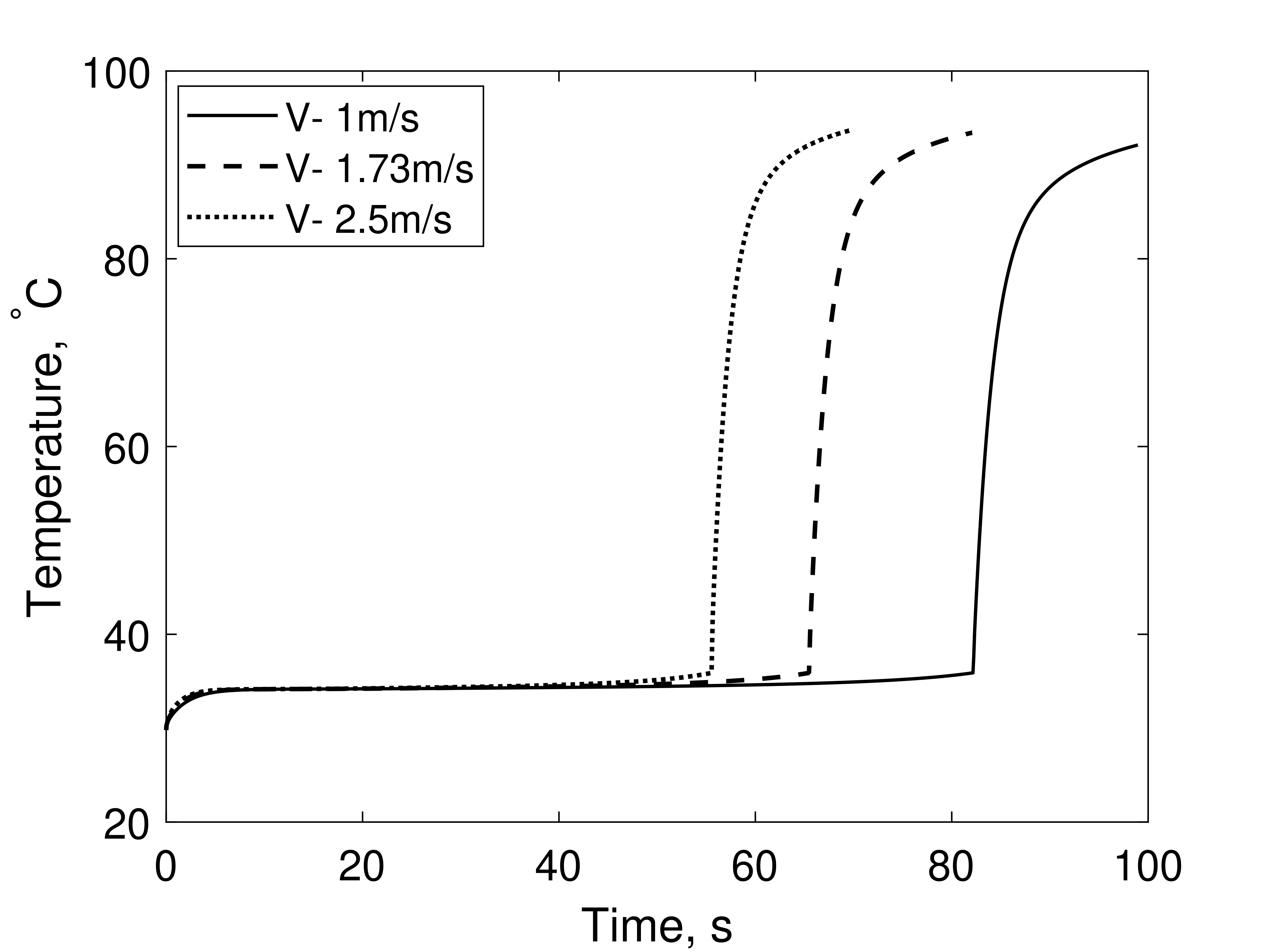}};
    \draw (-4, -6.5) node[inner sep=0] {\includegraphics[width=3.3in, trim=0cm 0cm 0cm 0cm, clip]{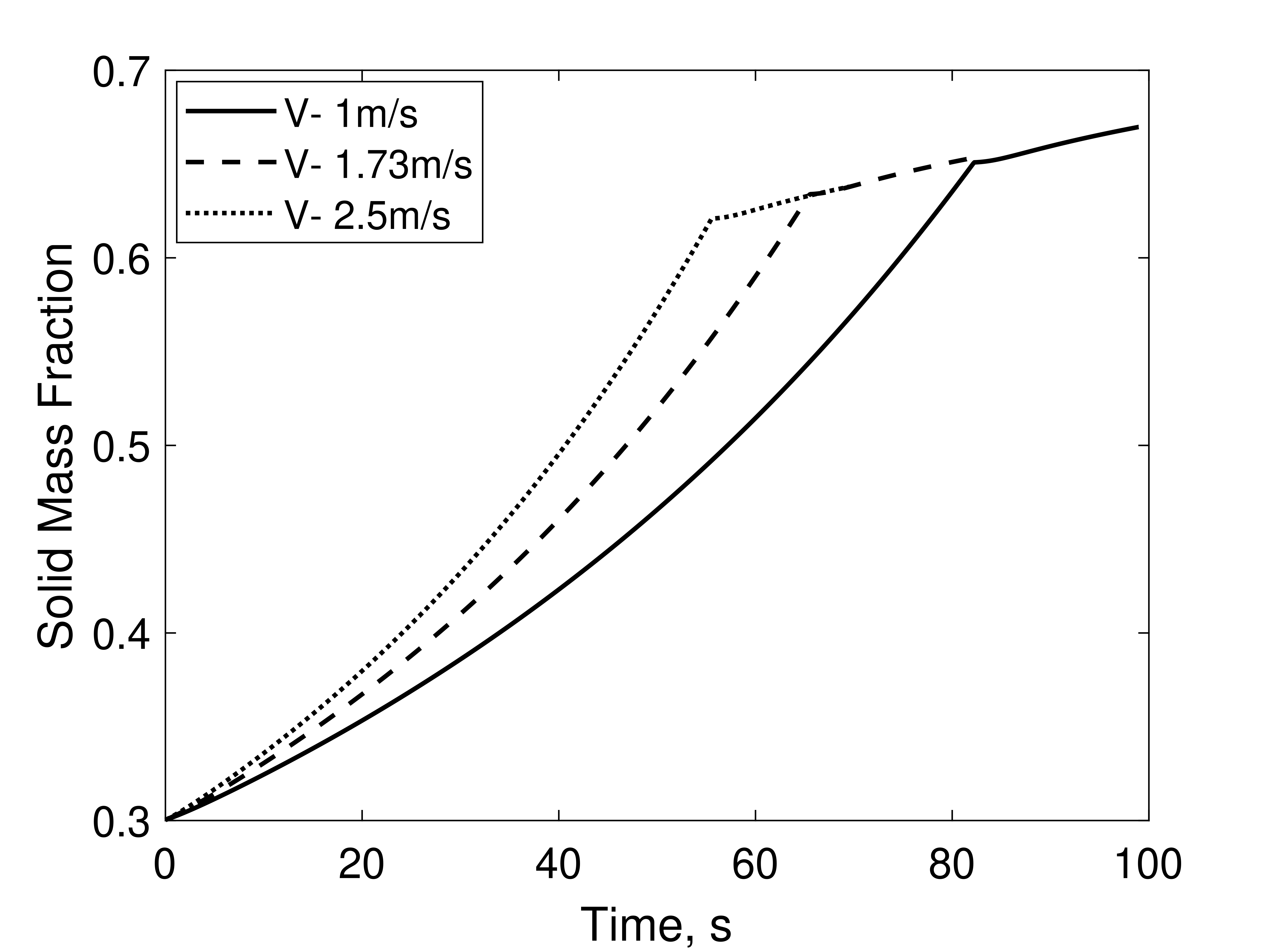}};
    \draw (4, -6.5) node[inner sep=0] {\includegraphics[width=3.3in, trim=0cm 0cm 0cm 0cm, clip]{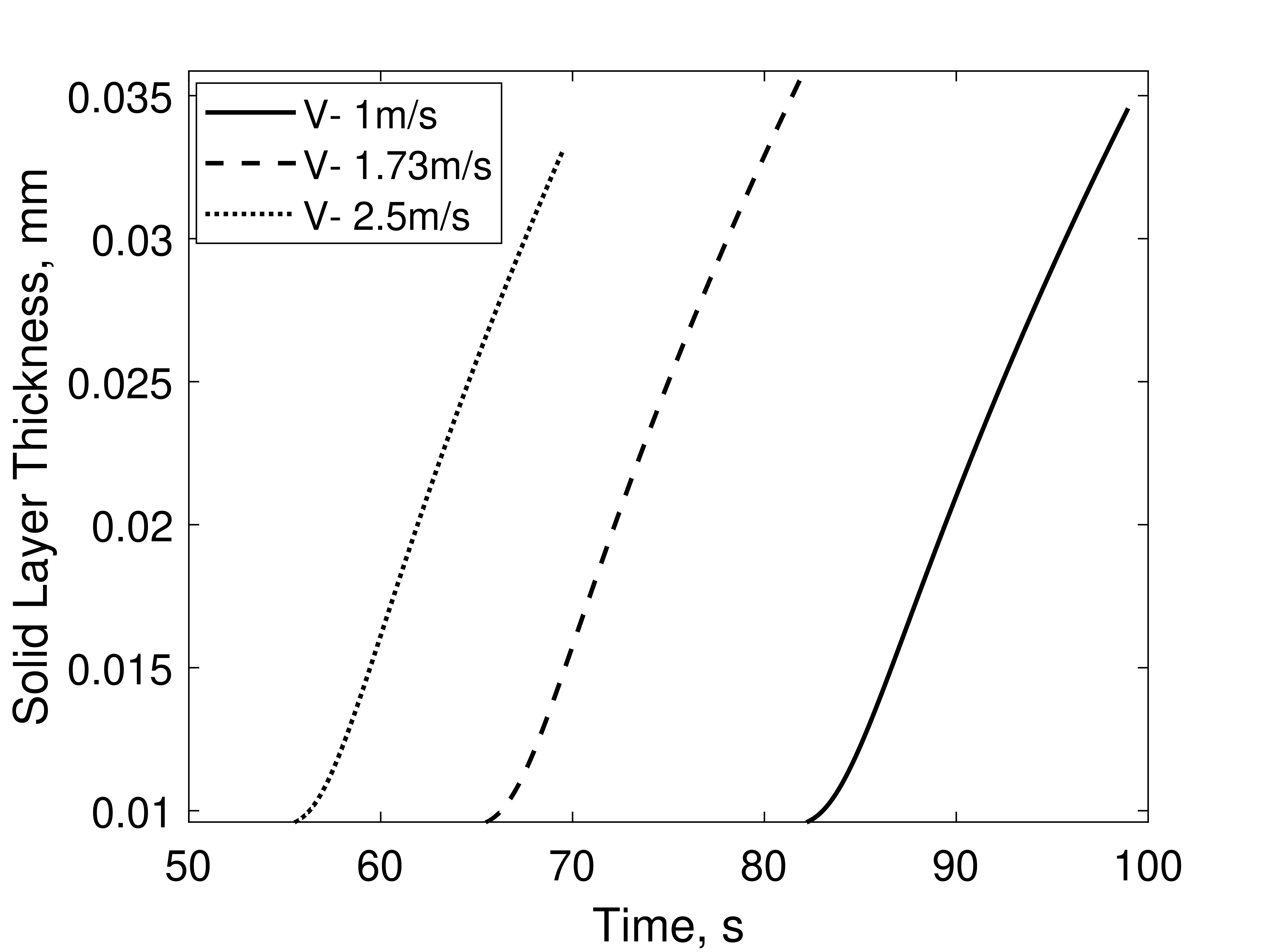}};    
    \draw (-7, 3.2) node[scale=1]{$(a)$};
    \draw (1, 3.2) node[scale=1]{$(b)$};
    \draw (-7, -3.3) node[scale=1]{$(c)$};
    \draw (1, -3.3) node[scale=1]{$(d)$};     
\end{tikzpicture}
\caption{Effect of drying gas velocity on drying of a  colloidal silica droplet. Figures show evolution of (a) mass (b) temperature (c) solid mass fraction and (d) solid layer thickness.}
\label{fig:velocity}
\end{figure}

The velocity of the droplets within the drying chamber can vary significantly during their travel. Initially, the droplets are injected with a high velocity near the atomizer, but they may reach a terminal velocity as they move down the chamber. Figure~\ref{fig:velocity} shows the temperature and mass of the droplets for three different velocities. It is evident that the drying time is greatly influenced by the velocity of the drying gas. Although the mass of the droplets changes considerably, the temperature evolution during the initial stage remains relatively unaffected. The mass evolution is expected to be more influenced, as the airflow over the droplet surface increases the rate of evaporation, leading to faster moisture loss. The velocity also affects the timing of the first stage, with faster crust formation occurring at higher velocities. Similar to the drying gas temperature, a higher gas velocity helps in the removal of moisture during the first stage, resulting in an increased solid mass fraction at higher drying gas velocities, as shown in Figure~\ref{fig:velocity}(c). However, it is interesting to note that the moisture removal rates for different gas velocities are similar during the second stage. This could be because the wet-core, where evaporation occurs, is not directly in contact with the drying gas due to the presence of a solid crust, which provides an additional layer of protection. As a result, both the solid mass fraction and crust thickness evolve at the same rate during the second stage, as highlighted in Figure~\ref{fig:velocity}(c) and (d).

\vspace{-0.1in}

\subsubsection{Development of a Regime Map for Predicting Particle Morphology}

\begin{figure}[H]
\begin{tikzpicture}[scale=1]
    \draw (-4, 0) node[inner sep=0] {\includegraphics[width=3.3in, trim=0cm 0cm 0cm 0cm, clip]{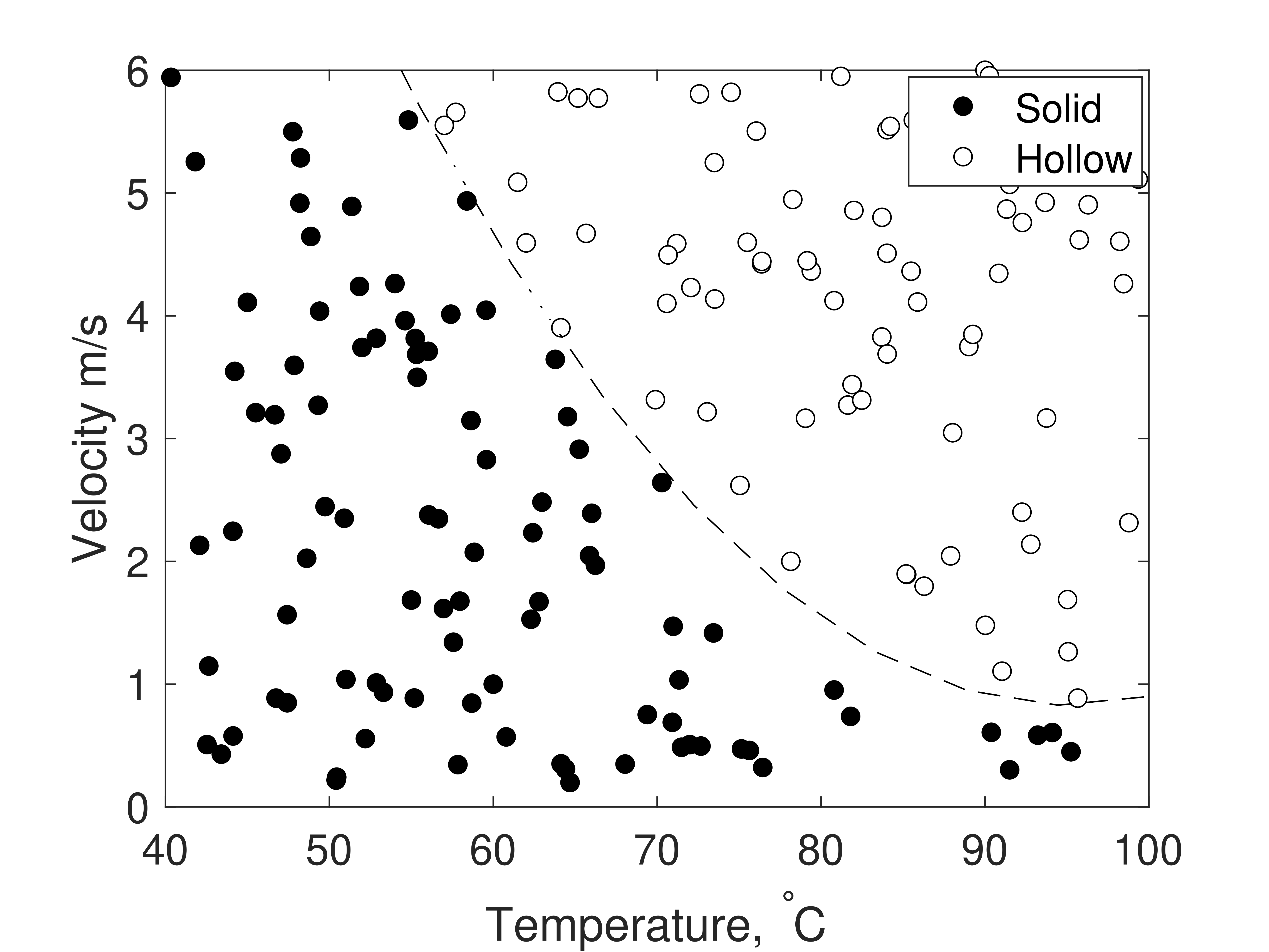}};
    \draw (4, 0.5) node[inner sep=0] {\includegraphics[width=3.4in, trim=0cm 0cm 0cm 0cm, clip]{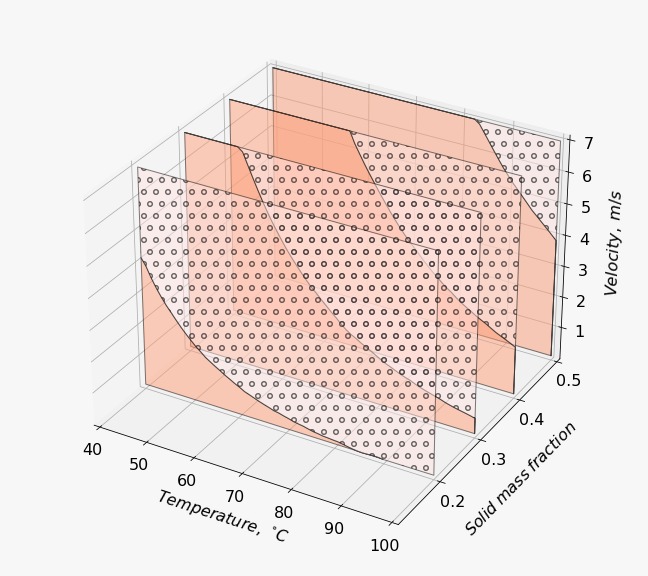}};
    \draw (-7, 3.2) node[scale=1]{$(a)$};
    \draw (1, 3.2) node[scale=1]{$(b)$};
\end{tikzpicture}
\caption{Regime map depicting solid and hollow particle formation at different drying conditions (a)  2-D plot for solid mass fraction of 0.3; (b) 3-D regime map at four different solid mass fractions.}
\label{fig:solid_hollow}
\end{figure}

For a constant porosity model that is considered in this study, it is possible to determine the final morphology of the particle from the properties at the end of the first stage. We present here the derivation of such a formula and develop a regime map that can predict whether the particle is solid or hollow under various drying conditions. Let $R_i$ be the radius of the interface at any instance of time during the second stage and $R_s$ be the outer radius of the particle, which does not change with time. Now, porosity, $\epsilon$ can be defined as:
\begin{equation}
    \epsilon = 1 - \frac{V_{sol}}{V_{crust}},
\end{equation}
where $V_{sol}$ and $V_{crust}$ refer to the volume of crust occupied by solid and total volume of crust respectively. From this we get,
\begin{equation}
    V_{sol} = V_{crust}\times(1-\epsilon).
\end{equation}
From the volume of solid in the crust, we can determine the mass of solid in the crust, using following relation:
\begin{equation}
    M_{sol} = V_{crust}\rho_{sol}(1-\epsilon),
\end{equation}
where $M_{sol}$ and $\rho_{sol}$ represent the mass of solid in the crust and density of the solid particles respectively. At the end of the drying when all the moisture has evaporated from the wet-core, the total mass of solid in the crust should be equal to the total mass of solid in the droplet, $M_s$:
\begin{equation}
    M_s=\frac{4}{3}\pi (R_s^3 - R_i^3)\times (1-\epsilon)\rho_{sol}.
\end{equation}
The above equation can be rewritten as:
\begin{equation}\label{eqn:solid_hollow}
    R_s = \sqrt[3]{R_i^3 + \frac{3 M_s}{4\pi (1-\epsilon)\rho_s} }.
\end{equation}
 It can be inferred that the final dried particle will be hollow if $R_i$ is greater than zero at the end of drying. Therefore, it can be concluded that to obtain a hollow particle at the end of drying, the following relation must be satisfied:
\begin{equation}
    R_s > \sqrt[3]{\frac{3 M_s}{4\pi (1-\epsilon)\rho_s} }.
\end{equation}
The end result will be a solid particle if
\begin{equation}\
    R_s \leq \sqrt[3]{\frac{3 M_s}{4\pi (1-\epsilon)\rho_s} }.
\end{equation}
By varying the temperature and velocity of the drying gas, we can create a regime map that illustrates the final particle morphology. Figure~\ref{fig:solid_hollow}(a) displays a regime map that depicts the formation of solid and hollow particles for a specific solid mass fraction. To construct a 3D regime map, we repeat this process for different solid mass fractions, as shown in Figure~\ref{fig:solid_hollow}(b). Hollow particles are formed when the moisture removal rate is significantly faster than the solid diffusivity towards the center of the droplet. This phenomenon is evident in Figures~\ref{fig:solid_hollow}(a) and (b), where we observe solid particles at low temperatures and velocities. Under these conditions, the moisture removal rates are lower, allowing sufficient time for the solid to diffuse towards the center of the droplet. However, as the initial solid mass fraction increases, solid particles are more likely to form even at relatively higher temperatures and velocities, as there is a greater amount of solid component present in the droplet.


\section{Summary}\label{}

A numerical model is developed to study the drying of a slurry droplet containing suspended solid particles. The model is constructed based on dividing the droplet drying process into three distinct stages. In the first stage, the liquid evaporates from the surface while the solid particles diffuse in the droplet. In the second stage, a dry porous crust is assumed to form on the surface and liquid continues evaporating through the porous crust. The final stage is the sensible heating of the dried particle, once all the moisture has been evaporated. Heat and mass transfer equations are solved in spherical coordinates to model the droplet mass and temperature evolution. The results are validated against experimental data showing good agreement. 

To investigate the impact of different modeling parameters and assumptions on the drying process, a comprehensive parametric study is conducted. The results reveal that the initial thickness of the crust has minimal influence, while the shape of the solid particles has a notable effect on the temperature evolution of the droplet. Furthermore, the validity of assuming a continuum flow for vapor transport through the crust pores is evaluated, and it is found that accounting for non-continuum effects significantly alters the temperature evolution during the second stage of drying. Additionally, the suitability of a uniform temperature model is examined and found to be inadequate for Biot numbers exceeding 0.1.

To investigate the impact of the drying gas parameters on changes in droplet mass, temperature, and final morphology, another set of parametric studies is conducted. The results reveal that the velocity and temperature of the drying gas significantly influence the final droplet morphology, whereas the relative humidity of the drying gas has only a minor impact. A functional relationship is established to predict the final morphology of the particle and a regime map is constructed for various initial solid mass fractions. By utilizing the regime map, it becomes possible to predict whether the final particle will be solid or hollow for a given drying gas velocity and temperature.

\section*{Acknowledgment}
 This study was supported by the Center for Advanced Research in Drying (CARD), a US National Science Foundation Industry/University Cooperative Research Center. CARD is located at Worcester Polytechnic Institute and University of Illinois at Urbana-Champaign (co-site). We also thank Mr. Diego Vaca-Revelo for his help in creating the 3D regime map.

\newpage
\appendix
\section*{APPENDIX}
\renewcommand{\thesubsection}{\Alph{subsection}}
\subsection{Calculating Thermo-Physical Properties}

\subsubsection{Thermo-Physical Properties Of Air}
The calculation of external heat and mass transfer rates in the drying model relies on understanding the thermo-physical characteristics of air. Numerous correlations exist in literature for this purpose. Although a comprehensive critical review was not conducted, the formulae incorporated in the model, as outlined below, are sourced from recognised sources. The validity of many relationships has been independently confirmed through cross-referencing with original references or alternative data sources.

The following relations have been taken from literature~\cite{werner2005air}. Most of these relations have been established by Adhikari~\cite{adhikari2003drying,adhikari2002application}.
\subsubsection*{Thermal Conductivity}
The thermal conductivity of air is expressed as:
\begin{equation*}
    K_a = 1.97\times 10^{-4}(T_{a})^{0.858},
\end{equation*}
in $[W.m^{-1}.K^{-1}]$, valid in the range of 0 - 205 $^{\circ}$C. Here, $T_a$ is the temperature of air in $[K]$.

\subsubsection*{Specific Heat Capacity}
The specific heat of air can be expressed as a function of temperature using the following relation:
\begin{equation*}
    C_{p,a} = 4\times 10^{-4}(T_{a}-273.15)^2+2.38\times 10^{-2}(T_{a}-273.15)+1004.5,
\end{equation*}
in $[J.Kg^{-1}.K^{-1}]$ valid in the range of 0 - 205 $^{\circ}$C, taken from Adhikari~\cite{adhikari2003drying}. 

\subsubsection*{Density}
The density of air can be calculated from the following relation:
\begin{equation*}
    \rho_a = \frac{P_t MW_a}{R_g T_{a}},
\end{equation*}
in $[Kg.m^{-3}]$ using ideal gas law, where $P_t$, $MW_a$ and $R_g$ are ambient pressure, molar mass of air and universal gas constant respectively.

\subsubsection*{Viscosity}
Viscosity of air depends on the temperature and can be computed from the following equation-
\begin{equation*}
    \mu_a = 1.097\times 10^{-6}\frac{(T_{a})^{0.5}}{1.453 - 0.0243(T_{a})^{0.5}},
\end{equation*}
in $[Pa.s]$. The relation is obtained from Sano and Keey~\cite{sano1982drying}.

\subsubsection{Thermo-Physical Properties Of Water}

The computation of external heat and mass transfer rates in the drying model is contingent upon correctly calculating the thermo-physical properties of water. It is important to note that the solvent used in the slurry droplet in this study is water. Moreover, in spray drying applications, water is the most commonly used solvent. In literature, various correlations exist for this purpose.  The model incorporates formulas outlined below, sourced from recognized references, although a comprehensive critical review has not been undertaken. 

\subsubsection*{Latent Heat of Vaporization}
The latent heat of vaporization of water can be computed using the following relation:
\begin{equation*}
    h_{fg} = -0.0013(T_d-273.15)^2-2.29618(T_d-273.15)+2500,
\end{equation*}
in $[KJ.Kg^{-1}]$, where $T_d$ is the droplet temperature in $[K]$. This is derived using the least square technique, utilizing data from Rahman~\cite{rahman2009food} across a temperature range spanning 0 to 100 $^{\circ}$C.

\subsubsection*{Specific Heat Capacity}
The specific heat capacity of water can be expressed using:
\vspace{-0.07in}
\begin{multline*}
    C_{p,w} =  2.108052\times 10^{-9}(T_d)^4-2.841073\times 10^{-6}(T_d)^3+1.441786\times 10^{-3}(T_d)^2-\\
    3.260186\times 10^{-1}(T_d)+3.18591\times 10^{-1},
\end{multline*}

in $[KJ.Kg^{-1}.K^{-1}]$, taken from Aspen HYSYS V8.4~\cite{aspenhysys}.

\subsubsection*{Thermal Conductivity}
The thermal conductivity of water can be expressed as a function of droplet temperature using the following relation obtained from Aspen HYSYS V8.4~\cite{aspenhysys}.
\vspace{-0.07in}
\begin{multline*}
   K_{w} = 5.33818\times 10^{-10}(T_d)^4-6.91901\times 10^{-7}(T_d)^3+3.25465\times 10^{-4}(T_d)^2-\\
   6.44535\times 10^{-2}(T_d)+5.01189,
\end{multline*}
in $[W.m^{-1}.K^{-1}]$.

\subsubsection*{Dynamic Viscosity}
The dynamic viscosity of water is also a function of temperature and can be expressed as:
\vspace{-0.07in}
\begin{multline*}
    \mu_{w} = 2.338519\times 10^{-11}(T_d)^4-3.24421\times 10^{-8}(T_d)^3+1.692507\times 10^{-5}(T_d)^2- \\
    3.940708\times 10^{-3}(T_d)+3.463261\times 10^{-1},
\end{multline*}

in $[N.s.m^{-2}]$, obtained from Aspen HYSYS V8.4~\cite{aspenhysys}.

\subsubsection{Solute Properties}

The physical properties for the colloidal silica  that are used for the validation studies are listed as follows:

\begin{center}
\begin{tabular}{ |c|c| }
\hline
 Properties & Silica  \\ 
 \hline
 Density [$\frac{kg}{m^3}$] & 1800 \\  
 Thermal Conductivity [$\frac{W}{m.K}$] & 0.660  \\
 Specific Heat [$\frac{J}{Kg.K}$] & 700   \\
 Molar Mass [$\frac{Kg}{mol}$] & 0.06008  \\
 Porosity  & 0.4  \\
 \hline
\end{tabular}
\end{center}

\bibliographystyle{myunsrtnat}

\newpage
\bibliography{references.bib}

\end{document}